\documentclass[floats,floatfix,showpacs,amssymb,prd,twocolumn,superscriptaddress,nofootinbib,nolongbibliography,reprint]{revtex4-2}

\usepackage{amssymb,amsmath,verbatim,mathtools,needspace,enumitem,etoolbox,graphicx,physics,microtype,afterpage,bigints,gensymb,tabularx,mathtools}

\usepackage[dvipsnames, usenames]{xcolor}
\definecolor{linkcolor}{rgb}{0.0,0.3,0.5}
\definecolor{dodgerblue}{HTML}{1E90FF}
\usepackage[unicode, colorlinks=true, linkcolor=linkcolor, citecolor=linkcolor, filecolor=linkcolor,urlcolor=linkcolor, pdfusetitle]{hyperref}
\usepackage[all]{hypcap}
\usepackage[T1]{fontenc}
\usepackage[utf8]{inputenc}
\usepackage{orcidlink}

\makeatletter

\usepackage{scalerel}
\usepackage{stackengine,wasysym}

\usepackage{titlesec}
\titleformat{\subsubsection}{\centering\bfseries\itshape}{\thesubsubsection.}{1em}{}

\newcommand{\ssim}{\mathchar"5218\relax\,}

\renewcommand{\vec}[1]{\mathbf{#1}}
\DeclareMathOperator\sign{sgn}
\DeclareMathOperator\sn{sn}

\makeatletter
\newcommand*{\balancecolsandclearpage}{\close@column@grid \cleardoublepage \twocolumngrid}
\makeatother

\usepackage{lipsum}

\def\apj{{Astrophys.~J.}}\def\apjl{{Astrophys.~J.~Lett.}}

\def\mnras{{Mon.~Not.~R.~Astron.~Soc.}}

\def\prd{{Phys.~Rev.~D}}

\def\prx{{Phys.~Rev.~X}}
\def\prl{{Phys.~Rev.~Lett.}}

\def\nat{{Nature}}

\def\cqg{{Classical~Quant.~Grav.}}

\newcommand{\bham}{\affiliation{School of Physics and Astronomy \& Institute for Gravitational Wave Astronomy, University of Birmingham, \\ Birmingham, B15 2TT, United Kingdom}}

\newcommand{\milan}{\affiliation{Dipartimento di Fisica ``G. Occhialini'', Universit\'a degli Studi di Milano-Bicocca, Piazza della Scienza 3, 20126 Milano, Italy}}
\newcommand{\infn}{\affiliation{INFN, Sezione di Milano-Bicocca, Piazza della Scienza 3, 20126 Milano, Italy}}

\newcommand{\florida}{\affiliation{Department of Physics, Florida International University, Miami, Florida 33199, USA}}

\begin{document}

\title{Efficient multi-timescale dynamics of precessing black-hole binaries 
}

\author{Davide Gerosa$\,$\orcidlink{0000-0002-0933-3579}}
\email{davide.gerosa@unimib.it}
\milan \infn \bham

\author{Giulia Fumagalli$\,$\orcidlink{0009-0004-2044-989X}} 
\milan \infn

\author{Matthew Mould$\,$\orcidlink{0000-0001-5460-2910}}
\bham

\author{\\Giovanni Cavallotto$\,$\orcidlink{0009-0001-5914-0361}}
\milan

\author{Diego Padilla Monroy$\,$\orcidlink{0009-0006-7758-7980}}
\florida \milan

\author{Daria Gangardt$\,$\orcidlink{0000-0001-7747-689X}}
\bham

\author{Viola De Renzis$\,$\orcidlink{0000-0001-7038-735X}}
\milan \infn


\date{\today}

\begin{abstract}

We present analytical and numerical progress on black-hole binary spin precession at second post-Newtonian order using multi-timescale methods. In addition to the commonly used effective spin which acts as a constant of motion, we exploit the weighted spin difference and show that such reparametrization cures the coordinate singularity that affected the previous formulation for the case of equal-mass binaries. The dynamics on the precession timescale is written down in closed form in both coprecessing and inertial frames. Radiation reaction can then be introduced in a quasi-adiabatic fashion such that, at least for binaries on quasi-circular orbits, gravitational inspirals reduce to solving a single ordinary differential equation. We provide a broad review of the resulting phenomenology and rewrite the relevant physics in terms of the newly adopted parametrization. This includes the spin--orbit resonances, the up--down instability, spin propagation at past time infinity, and new precession estimators to be used in gravitational-wave astronomy. Our findings are implemented in version 2 of the public Python module {\sc precession}. Performing a precession-averaged post-Newtonian evolution from/to arbitrarily large separation takes $\lesssim 0.1$ s on a single off-the-shelf processor ---a $50\times$ speedup compared to our previous implementation. This allows for a wide variety of applications including propagating gravitational-wave posterior samples as well as population-synthesis predictions of astrophysical nature.
\end{abstract}

\maketitle

\section{Introduction}

Black-hole (BH) binary spin precession is a distinctive feature of the general relativistic two-body problem~\cite{1994PhRvD..49.6274A}. Two Kerr BHs in a bound system are subject to interactions between their spins and the orbital angular momentum of the binary. %
The motion resulting from such spin--orbit and spin--spin couplings is a superposition of precession and nutation.
This is somewhat analogous to that of Earth's axis, though with a key difference. For the Earth, precession happens on a much longer timescale ($\ssim 2.5 \times 10^4$ yr) than nutation ($\ssim 18$ yr) and with a larger amplitude, such that one can think of a toplike azimuthal motion perturbed by small polar oscillations. In the BH case, precession and nutation happen on comparable timescales and can have comparable amplitudes, leading to a more complex phenomenology. While the spins evolve, the system dissipates energy via gravitational-wave (GW) emission, ultimately leading to the merger of the two BHs. 

Timescale considerations are crucial to shed light on the dynamics of BH binaries, at least in the post-Newtonian (PN) regime. The orbital motion takes place on a timescale\footnote{Unless specified otherwise, we use natural units where $c=G=1$.} %
$t_{\rm orb} \propto (r/M)^{3/2}$ (by Kepler's law), the spins precess on $t_{\rm pre}\propto (r/M)^{5/2}$ \cite{1994PhRvD..49.6274A}, and radiation reaction takes place on $t_{\rm rad}\propto (r/M)^4$ (by the quadruple formula). At sufficiently large separations $r\gg M$ one has 
\begin{align}
\label{hierarchy}
t_{\rm orb} \ll t_{\rm pre} \ll t_{\rm rad}\,.
\end{align}
The first inequality $t_{\rm orb} \ll t_{\rm pre}, t_{\rm rad}$ has been used since the early foundation of GW physics~\cite{1963PhRv..131..435P}, resulting in the popular orbit-averaged formulation of the BH dynamics: sources evolve on quasi-Newtonian orbits and the orbit itself evolves quasi-adiabatically. More recently \cite{2015PhRvL.114h1103K,2015PhRvD..92f4016G},
some of us started investigating the consequences of the second inequality $t_{\rm pre}\ll t_{\rm rad}$.
The strategy is the same: we compute the ``shape'' of the precession cones on the short timescale (in this case $t_{\rm pre}$) and then implement radiation reaction in a quasi-adiabatic fashion. This approach allows for
extremely efficient evolutions of
BH binaries along their long inspirals before merger, which we first implemented in the numerical code \textsc{precession}~\cite{2016PhRvD..93l4066G}. When studying BH-binary inspirals, one can now time step over the longer timescale $t_{\rm rad}$ of the problem while the dynamics on both $t_{\rm orb}$ and $t_{\rm pre}$ is dealt with analytically. This is, at present, the only feasible strategy to evolve BH binaries from the (infinitely) large separations where they form down to the smaller separations where they enter the sensitivity windows of GW detectors.

Such a ``multi-timescale'' approach to the binary dynamics led to an explosion of new predictions and applications. This includes investigations on the phenomenology of BH binaries~\cite{
2015PhRvL.115n1102G,
2017CQGra..34f4004G,
2017PhRvD..96b4007Z,
2019CQGra..36j5003G,
2019PhRvD.100l4008P,
2020PhRvD.101l4037M,
2020CQGra..37v5005R,
2021PhRvD.103f4003V,
2021PhRvD.103l4026G,
2022PhRvD.106b3001J}, 
waveform development~\cite{
2017PhRvL.118e1101C,
2017PhRvD..95j4004C,
2019PhRvD.100b4059K,
2020PhRvD.102f4001P,
2021arXiv210610291K,
2022arXiv221204657L}, 
astrophysical modeling~
\cite{2015MNRAS.451.3941G,
2016ApJ...832L...2R,
2018PhRvD..98h4036G,
2019PhRvD..99j3004G,
2021MNRAS.501.2531S,
2021PhRvD.103f3032S,
2021PhRvD.104h4002B,
2022PhRvD.106f3028S,
2023MNRAS.519.5031S}, and 
 interpretation of current GW data~\cite{
2018PhRvD..97d3014W,
2018PhRvD..98h3014A,
2021PhRvD.103f4067G,
2022PhRvD.105b4076M,
2022PhRvD.106b4019G,
2022PhRvD.106h4040D,
2022CQGra..39l5003H,
2023arXiv230110125J}.
The \textsc{precession} code itself~\cite{2016PhRvD..93l4066G} was used in
over 60
publications to date.
Most notably, (i) the analytic treatment of the binary dynamics on the precession timescale is now part of the standard ``twisting up'' strategy used in state-of-the-art phenomenological waveform models \cite{2019PhRvD.100b4059K,2020PhRvD.102f4001P}, and (ii) the latest GW catalogs and population analyses developed by LIGO/Virgo now quote spin directions that are back-propagated to infinitely large separations using precession-averaged equations~\cite{2021arXiv211103606T,2023PhRvX..13a1048A}.

In this paper, we present a full reinvestigation of BH-binary spin precession using multi-timescale methods. We exploit and expand upon some recent analytical advances~\cite{2021arXiv210610291K} which allow us to write the entire dynamics on the precession timescale $t_{\rm pre}$ in closed form
using Legendre elliptic integrals and Jacobi elliptic functions (Sec.~\ref{precdyn}). This new formulation cures a coordinate singularity that impacted the previous parametrization for equal-mass BHs \cite{2017CQGra..34f4004G,2022PhRvD.106b3001J}.
While the dynamics on the radiation-reaction timescale still needs to be integrated numerically (Sec.~\ref{precavinspiral}), this operation is
$\gtrsim 50$ times faster compared to our previous implementation \cite{2016PhRvD..93l4066G} ---BH binaries can now be evolved all the way from/to past time infinity in $\lesssim 0.1$ s on a single processor. We then review the broader phenomenology of spin precession in BH binaries and rewrite the relevant equations in terms of the new regularized quantities (Sec.~\ref{phenosec}). We provide some additional ingredients that are often necessary to interface the precession-averaged formalism with other results in BH physics (Sec.~\ref{interface}). Our findings are implemented in v2 of the \textsc{precession} code, which has been rewritten from scratch. The code is publicly available at \href{https://github.com/dgerosa/precession}{github.com/dgerosa/precession}~\cite{repo}, where we also provide documentation, tutorials, and various PN coefficients in machine-readable format. Here we briefly report on the performance of the new code as well as some profiling results (Sec.~\ref{pythonsec}). We conclude with a roadmap for the future development of the precession-averaged approach to the BH binary spin precession problem (Sec.~\ref{concl}). Lengthy PN expression, details on our algorithm, and some mathematical expressions are postponed to the Appendixes.

\section{Precession dynamics}
\label{precdyn}

\subsection{Looking for an optimal parametrization}

\label{lookfor}
 
Let us consider a BH binary with masses $m_{1}\geq m_2$, orbital angular momentum $\boldsymbol{L}$, and spins $\boldsymbol{S}_{1,2}$.  These are combined into the total  spin $\boldsymbol{S} =\boldsymbol{S}_1 + \boldsymbol{S}_2$ and the total angular momentum $\boldsymbol{J} = \boldsymbol{L} + \boldsymbol{S}_1 + \boldsymbol{S}_2$.  The total mass $M=m_1+m_2$ is a free scale of the problem and can thus be treated as a unit (indeed, in our numerical implementation we simply set $M=1$ and measure all other quantities accordingly; cf. Sec.~\ref{pythonsec}). We then define the mass ratio $q=m_2/m_1\in(0, 1]$ and the Kerr parameters $\chi_{1,2}=S_i/m_i^2 \in [0,1]$ such that the spin magnitudes are given by
\begin{align}
\label{spinmags}
S_1 &= \frac{\chi_1}{(1+q)^2} M^2\,,
\\
S_2 &= \frac{q^2\chi_2}{(1+q)^2} M^2\,.
\end{align}
We restrict to sources on quasi-circular orbits; a generalization to eccentric orbits is under active development and will be presented elsewhere. The magnitude of the (Newtonian) orbital angular momentum can be expressed in terms of either the orbital separation $r$, via
\begin{align}
L= M^2 \frac{q}{(1+q)^2}\sqrt{\frac{r}{M}}\,,
\end{align}
or the compactified coordinate~\cite{2015PhRvD..92f4016G} %
\begin{align}
\label{udef}
u = \frac{1}{2L} = \frac{(1+q)^2}{2 q M^2} \sqrt{\frac{M}{r}}\,,
\end{align}
such that $u\to 0$ as $r\to \infty$.

In a frame that co-precesses with the binary \cite{2011PhRvD..84b4046S,2011PhRvD..84l4002O,2011PhRvD..84l4011B}, the mutual orientations of $\boldsymbol{L}$, $\boldsymbol{S}_{1}$ and $\boldsymbol{S}_2$ are fully described by three angles~\cite{2004PhRvD..70l4020S}. These are often chosen to be the polar angles $\theta_{1,2}\in [0,\pi]$ between the spin and orbital angular momentum,
\begin{align}
\label{costheta1}
\cos\theta_1 &=\hat{\boldsymbol{S}}_1 \cdot \hat{\boldsymbol{L}}\,,
\\
\cos\theta_2&=\hat{\boldsymbol{S}}_2 \cdot \hat{\boldsymbol{L}}\,,
\label{costheta2}
\end{align}
(where a hat denotes a unit vector) %
and the azimuthal angle $\Delta\Phi\in [-\pi,\pi]$ between the projections of the two spins onto the orbital plane,\footnote{Other works in the literature use the symbol $\phi_{12}$ for $\Delta\Phi$.}
\begin{align}
\label{cosdphi}
\cos\Delta\Phi&=\frac{\hat{\boldsymbol{S}}_1 \times \hat{\boldsymbol{L}}}{|\hat{\boldsymbol{S}}_1 \times  \hat{\boldsymbol{L}} |} \cdot 
\frac{\hat{\boldsymbol{S}}_2 \times \hat{\boldsymbol{L}}}{|\hat{\boldsymbol{S}}_2 \times \hat{\boldsymbol{L}} |},
\\
\sign\Delta\Phi &= \sign\{ \boldsymbol{L} \cdot [(\boldsymbol{S}_1 \times \boldsymbol{L}) \times (\boldsymbol{S}_2 \times \boldsymbol{L})] \}.
\label{dphisign}
\end{align}
From these, one can obtain the angle $\theta_{12}$ between the two spins within the plane they generate:
\begin{align}
\cos\theta_{12} = \hat{\boldsymbol{S}}_1 \cdot \hat{\boldsymbol{S}}_2 = \cos\theta_1\cos\theta_2 +\cos \Delta\Phi \sin\theta_1\sin\theta_2\,.
\label{costheta12angles}
\end{align}

Following the notation first introduced in Ref.~\cite{2015PhRvL.115n1102G}, we refer to spinning but non-precessing binaries as ``up--up'' for $\theta_1=\theta_2=0$, ``up--down'' for $\theta_1=0$ and $\theta_2=\pi$, ``down--up'' for $\theta_1=\pi$ and $\theta_2=0$, and ``down--down'' for $\theta_1=\theta_2=\pi$.

While intuitive, parametrizing the spin evolution using $\theta_1$, $\theta_2$, and $\Delta\Phi$ significantly complicates the dynamics because all three angles vary on the same timescale $t_{\rm pre}$. Instead, the evolution of GW sources can be greatly simplified by identifying quantities that respect the timescale hierarchy of Eq.~(\ref{hierarchy}).
\citeauthor{2008PhRvD..78d4021R}~\cite{2008PhRvD..78d4021R} first realized that the effective spin \cite{2001PhRvD..64l4013D}
\begin{align}
\chi_{\rm eff}= \frac{\chi_1 \cos\theta_1 + q \chi_2\cos\theta_2}{1+q}
\label{chieff}
\end{align}
is a constant of motion at 2PN. This is also the spin quantity that is commonly regarded as best measured from LIGO/Virgo observations \cite{2019PhRvX...9c1040A,2021PhRvX..11b1053A,2021arXiv211103606T}.
Both the separation $r$ and the magnitude of the total angular momentum 
\begin{align}
J 
& = [L^2+ S_1^2 + S_2^2 + 2L S_1 \cos\theta_1 
 + 2 L S_2 \cos\theta_2 
 \notag \\
&+2  S_1 S_2 (\cos\theta_1\cos\theta_2 +\cos \Delta\Phi \sin\theta_1\sin\theta_2)]^{1/2} \end{align}
change solely because of GW emission. These parameters are therefore constant on the precession timescale and vary only on the radiation-reaction timescale. The direction $\hat{\boldsymbol{J}}$ is approximately constant even on this longer timescale, with the notable exception of cases where the magnitude $J$ approaches zero \cite{1994PhRvD..49.6274A}; see also Ref.~\cite{2017PhRvD..96b4007Z} for a quantitative analysis on this point.

With these considerations, some of the authors \cite{2015PhRvL.114h1103K,2015PhRvD..92f4016G} realized that entire dynamics on the precession timescale can be reduced to the evolution of a single quantity. This is analogous to the effective potentials in Kepler's two-body problem, where energy and angular momentum conservation reduce the motion to that of an equivalent particle in one dimension. In particular, previous work parametrizes BH-binary spin precession using the magnitude of the total spin 
\begin{align}
\label{Sgeodef}
S 
&= [S_1^2 \!+\! S_2^2 \!+ \! 2S_1 S_2 (\cos\theta_1\!\cos\theta_2 \!+\! \cos \Delta\Phi \!\sin\theta_1\!\sin\theta_2)]^{1/2}\!\!.
\end{align}
Using $(\chi_{\rm eff}, J, S)$ instead of $(\theta_1,\theta_2,\Delta\Phi)$ reflects the natural separation of timescale that governs the BH binary dynamics in the PN regime.

However, this parametrization is still suboptimal for two reasons. First, the magnitude of the total angular momentum $J$ diverges at large separations $r\to \infty$, making a numerical implementation impractical. This can be cured by instead using  %
\begin{align}
\label{kappadef}
\kappa  =\frac{J^2 - L^2}{2L}
\end{align}
which was proved to converge in the large separation limit~\cite{2015PhRvD..92f4016G}. We refer to $\kappa$ as the ``asymptotic angular momentum.'' The precise expression of Eq.~(\ref{kappadef}) has been chosen such that $\kappa$ reduces to $\boldsymbol{S}\cdot  \hat{\boldsymbol{L}} $ in the large-separation limit, see Sec.~\ref{notable}.

\begin{figure}
\includegraphics[width=\columnwidth]{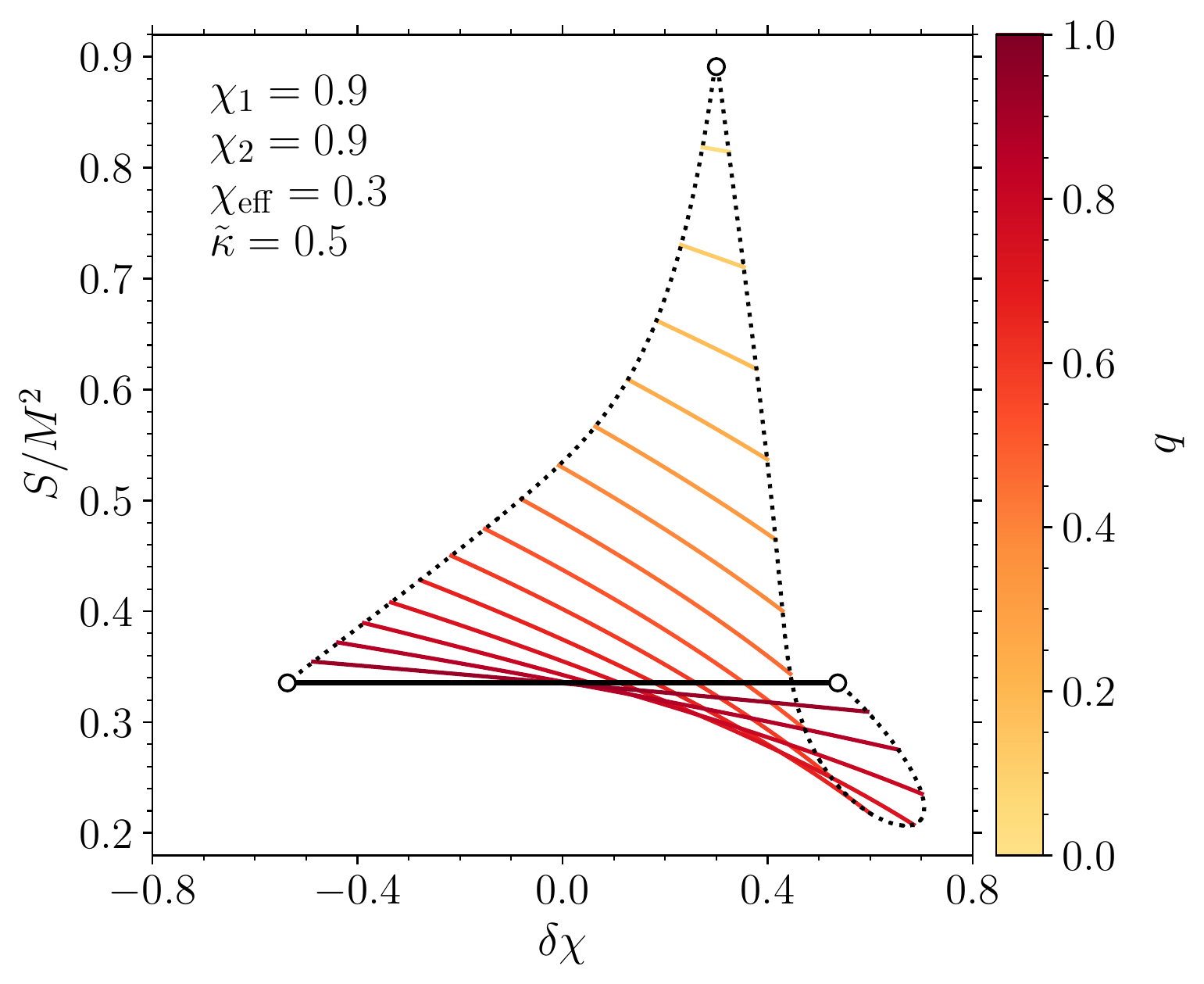}
\caption{Magnitude of the total spin $S$ as a function of the weighted spin difference $\delta\chi$. We consider a set of binaries with $\chi_1=\chi_2=0.9$, $\chi_{\rm eff}=0.3$, $\tilde\kappa=0.5$, and a range of values of $q\in(0,1]$ as indicated in the color bar. The dotted black curves mark the locations of the two turning points $\delta\chi_{\pm}$. The solid black line indicates the $q=1$ limiting case where the $S$ parametrization becomes degenerate. The $\delta\chi_{\pm}$ configurations  with $q=0,1$ are indicated with round markers.}
\label{Svsdeltachifig}
\end{figure}

Second, the magnitude of the total spin $S$ is a constant of motion for equal-mass binaries \cite{2008PhRvD..78d4021R}, which implies one cannot rely on $S$ to parametrize the precession cycle when $q=1$ \cite{2017CQGra..34f4004G}. This mathematical quirk is analogous to that of a coordinate singularity in general relativity, which does not affect the underlying physics but breaks the formalism. As we explore at length in this paper, the quantity
\begin{align}
\delta\chi= %
 \frac{\chi_1 \cos\theta_1 - q \chi_2\cos\theta_2}{1+q}
\label{deltachi}
\end{align}
first identified by \citeauthor{2021arXiv210610291K} \cite{2021arXiv210610291K} correctly regularizes the $q\to1$ limit. We refer to $\delta\chi$ as ``weighted spin difference.'' An alternative, approximate regularization scheme was recently proposed in Ref.~\cite{2022PhRvD.106b3001J}.
While Eqs.~(\ref{chieff}) and (\ref{deltachi}) are formally similar and only differ by a single sign, the dynamical properties of $\chi_{\rm eff}$ and $\delta\chi$ are crucially different. The former is a constant of motion for the 2PN spin problem while the latter varies on the smaller precession timescale. The conversion between the $S$ and $\delta\chi$ parametrizations is given by
\begin{align}
\frac{S^2}{M^4} =    \frac{q}{(1+q)^2}\sqrt{\frac{r}{M}} \left(2 \frac{\kappa}{M^2} - \chi_{\rm eff} -  \frac{1-q}{1+q}   \delta\chi \right)\,.
\label{Svsdeltachi}
\end{align}
Indeed, the precession-timescale variation encoded by $\delta\chi$ disappears for $q\to1$ such that $S$ tends to a constant. This is shown in  Fig.~\ref{Svsdeltachifig} for a representative set of sources: for $q=1$, the curve $S(\delta\chi)$ becomes horizontal, signaling the breakdown of the $S$ parametrization. %

We will now use Eq. (\ref{Svsdeltachi}) to rewrite and expand upon the entire formalism of Refs.~\cite{2015PhRvL.114h1103K,2015PhRvD..92f4016G}, thus regularizing the $q=1$ behavior.
The explicit expressions for the spin angles in terms of $\chi_{\rm eff}$, $\kappa$, and $\delta\chi$ are given by
\begin{align}
\label{ct1}
\cos\theta_1= &\frac{1+q}{2
 \chi_1} (\chi_{\rm eff}+\delta\chi)\,,
\\
\cos\theta_2 =  &\frac{1+q}{2 q
 \chi_2} (\chi_{\rm eff}-\delta\chi)\,,
  \end{align}
 \begin{align}
   \cos\Delta\Phi =&\frac{1}{q} \Big\{
 2 q (1 + q)\sqrt{r/M}  [2 (1 + q)  \kappa M^{-2}   
 \notag\\&
 -(1 + q)  \chi_{\rm eff}  
 - 
 (1 - q)     \delta\chi  ]  - 2 (\chi_1^2 + \chi_2^2  q^4) 
 \notag\\&
  -q (1 + q)^2 (\chi_{\rm eff}^2 - \delta\chi ^2) \Big\}
 \notag\\&
 \times
  \left[{4
   \chi_1^2-(1+q)^2 (\chi_{\rm eff}+\delta\chi)^2}\right]^{-1/2}
 \notag\\&
 \times   \left[{4 \chi_2^2 q^2-(1+q)^2
   (\chi_{\rm eff}-\delta\chi)^2}\right]^{-1/2}
   \,,
\label{cosdeltaphi} 
    \\
\cos\theta_{12} &= \frac{1}{2 \chi_1 \chi_2 q^2}
 \Big\{q (1 + q)\sqrt{r/M} [2  (1 + q) \kappa M^{-2}
\notag \\ 
 &- (1 + q)\chi_{\rm eff}   - 
   (1 - q) \delta\chi ]  - (\chi_1^2 + \chi_2^2  q^4) \Big\}
   \label{costheta12} \,.
   \end{align}
 Note that, unlike the analogous Eqs.~(20) in Ref.~\cite{2015PhRvD..92f4016G}, these equations are manifestly finite when $q=1$.

In summary, we parametrize BH binaries on quasi-circular orbits using $q$, $\chi_1$, $\chi_2$, $\chi_{\rm eff}$, $u$, $\kappa$, and $\delta\chi$. In particular:
\begin{itemize}
\item The mass ratio $q$, the spin magnitudes $\chi_{1,2}$ \cite{2018CQGra..35b4001I}, and the effective spin $\chi_{\rm eff}$ \cite{2008PhRvD..78d4021R} are all constants of motion at the PN order we consider.
\item The compactified orbital separation $u$ and the asymptotic angular momentum  $\kappa$ vary only on the radiation-reaction timescale and are asymptotically regular at large separations.
\item The weighted spin difference $\delta\chi$ varies on the precessional timescale and is regular in the equal-mass limit.
\end{itemize}

\subsection{Dynamics in a co-precessing frame}
\label{rootssubsec}

\begin{figure*}
\includegraphics[width=\textwidth]{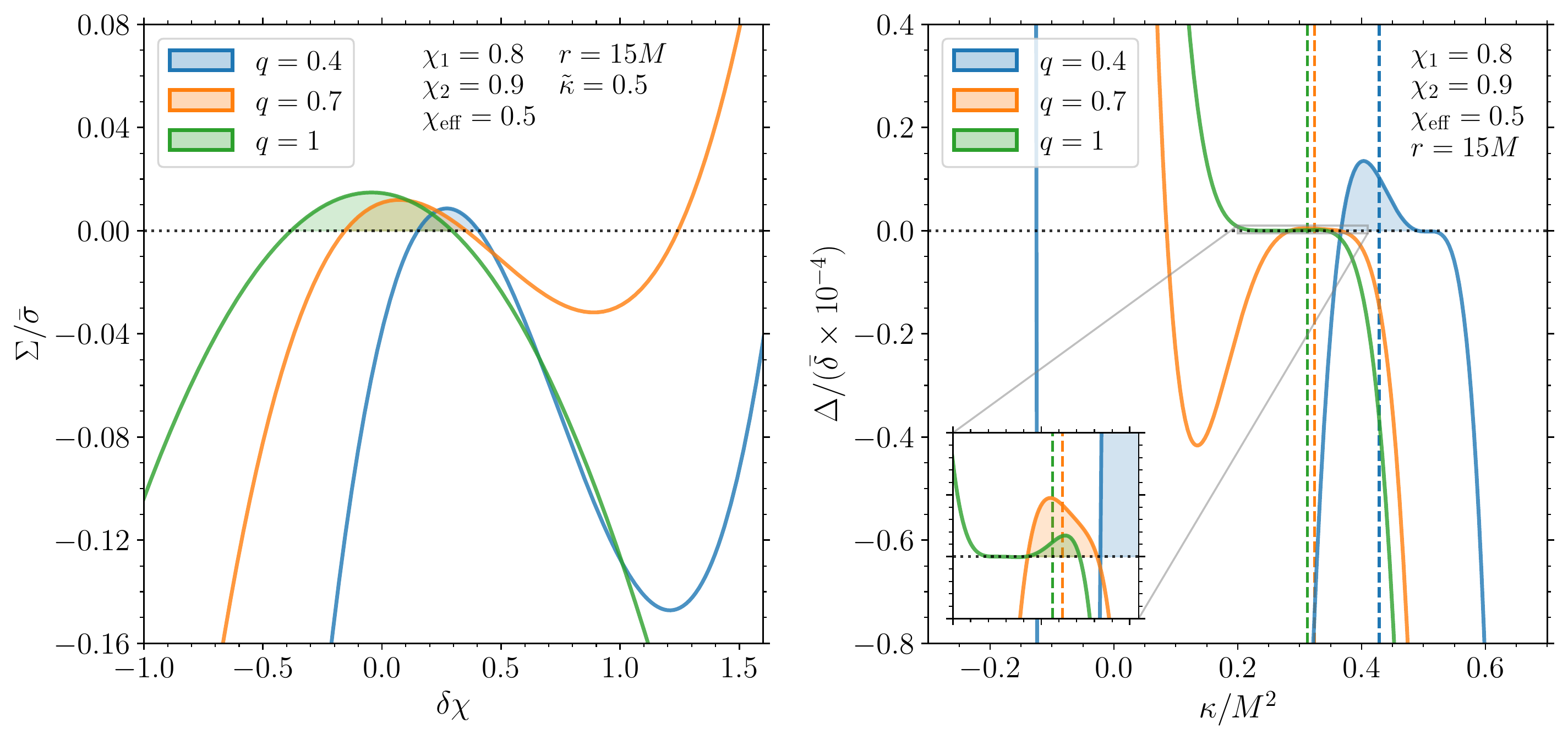}
\caption{The left panel shows the normalized cubic equation $\Sigma/\bar \sigma$ for the derivative $(\dd \delta\chi/\dd t)^2$ as reported in Eq.~(\ref{thirddegree}). Spin precession takes place in the bounded and positive intervals highlighted with the shaded areas. The left and right edges of these intervals correspond to $\delta\chi_-$  and $\delta\chi_+$, respectively. The spurious, largest root (which is absent for $q=1$) corresponds to $\delta\chi_3/(1-q)$. The right panel shows the discriminant $\Delta$ of the $\Sigma/\bar \sigma$ equation normalized to a positive coefficient $\bar \delta$; see Eq.~(\ref{deltakappa}). Spin precession can only take place in the bounded, positive intervals highlighted with the shaded areas. The edges of these intervals are the spin--orbit resonances $\kappa_\pm$.
In both panels, we consider binaries with $\chi_1=0.9$, $\chi_2=0.9$, $\chi_{\rm eff}=0.5$, $r=15M$, and three values of the mass ratio $q=0.4$ (blue), 0.7 (orange), and 1 (green). For the left panel, we further impose $\tilde\kappa=0.5$. The corresponding values of $\kappa$ are indicated with dashed lines in the right panel.
 }
\label{poly}
\end{figure*}

The evolution of the  spins and the orbital angular momentum is set by the coupled precession equations~\cite{1995PhRvD..52..821K,2008PhRvD..78d4021R,2006PhRvD..74j4033F,2006PhRvD..74j4034B,2010PhRvD..81h4054K}

\begin{align}
\label{dS1st}
\frac{\dd \boldsymbol{S}_1}{\dd t} &=  \boldsymbol{\omega}_1 \times \boldsymbol{S}_1\,,
\\
\frac{\dd \boldsymbol{S}_2}{\dd t} &=  \boldsymbol{\omega}_2\times \boldsymbol{S}_2\,,
\\
\frac{\dd \boldsymbol{L}}{\dd t} &=  \boldsymbol{\omega}_L\times \boldsymbol{L}  + \frac{\dd L}{\dd t} \hat{\vec{L}}\,,
\label{dLdtorb}
\end{align}
where the frequencies at 2PN are given by
\begin{align}
\label{spinomega1}
\boldsymbol{\omega}_1 &= \frac{1}{2 r^3}\left\{ \left[ 4+3q  - \frac{3 (1+q)\chi_{\rm eff}}{\sqrt{r/M}}\right]  \boldsymbol{L} + \boldsymbol{S}_2\right\} \,,
\\
\label{spinomega2}
\boldsymbol{\omega}_2 &= \frac{1}{2 r^3}\left\{ \left[ 4+\frac{3}{q}  - \frac{3 (1+q)\chi_{\rm eff}}{q\sqrt{r/M}}\right]  \boldsymbol{L} + \boldsymbol{S}_1\right\} \,,
\\ \nonumber
\boldsymbol{\omega}_L &= \frac{1}{2 r^3}\Bigg\{ \left[ 4+3q  - \frac{3 (1+q)\chi_{\rm eff}}{\sqrt{r/M}}\right]  \boldsymbol{S}_1 
\\ &+\left[ 4+\frac{3}{q}  - \frac{3 (1+q)\chi_{\rm eff}}{q\sqrt{r/M}}\right] \boldsymbol{S}_2\Bigg\} \,.
\label{OmegaL}
\end{align}
The derivative $\dd L /\dd t$ models radiation reaction and can be neglected when studying the dynamics on the precession timescale. In this approximation, the motion of a binary in a non-inertial, co-precessing frame is entirely encoded in the evolutionary equation of $\delta\chi$. From Eqs.~(\ref{dS1st})-(\ref{OmegaL}) one gets
\begin{align}
\label{ddeltachidt}
M \frac{\dd\delta \chi }{\dd t} &= \frac{3 q }{(1+q)^2}  \chi_1\chi_2 \left(\frac{r}{M}\right)^{-3} \left( 1 - \frac{\chi_{\rm eff}}{\sqrt{r/M}}\right) 
\notag \\
& \times \sin\theta_1 \sin\theta_2 \sin\Delta\Phi\,,
\end{align}
where the prefactor on the first line of Eq.~(\ref{ddeltachidt}) is always positive because $|\chi_{\rm eff}|\leq 1$ and $r>M$. In particular, the squared time derivative of $\delta\chi$ is given by 
\begin{align}
\label{thirddegree}
\Sigma (\delta\chi) \equiv \left(M \frac{\dd\delta \chi }{\dd t} \right)^2 = \bar\sigma (\sigma_3 \delta\chi^3 + \sigma_2 \delta\chi^2 +  \sigma_1 \delta\chi + \sigma_0)\,,
\end{align}
which is a third-degree polynomial. Some representative examples are shown in the left panel of Fig.~\ref{poly}.  The coefficients $\bar\sigma$ and $\sigma_i$ are functions of $q,\chi_1,\chi_2,\chi_{\rm eff},\kappa$, and $r$, as reported in Appendix~\ref{polycoeffs}. In particular, factorizing the equation in this form allows us to avoid divisions by zero in both the equal-mass ($q\to 1$) and large-separation ($u\to 0$) limits; cf. Sec.~\ref{notable}. 

The equation $\Sigma(\delta\chi)=0$ admits either one or three real roots.
Spin precession requires the existence of two turning points in the evolution $\delta\chi(t)$; a formal proof using the Jordan curve theorem is provided in Ref.~\cite{2015PhRvD..92f4016G}. This implies that, for physical configurations to exist, the equation $\Sigma (\delta\chi)=0$ must admit three roots, with two of them acting as the turning points that define spin precession. The third root is spurious and was introduced when squaring the derivative ${d\,\delta \chi }/{d t}$ to obtain Eq.~(\ref{thirddegree}) from Eq.~(\ref{ddeltachidt}).
In particular, it is useful to write~\cite{2021arXiv210610291K}
\begin{align}
\label{cubicandroots}
\left(\! M \frac{\dd\delta \chi }{\dd t} \right)^2 \!= \mathcal{A}^2 (\delta\chi - \delta\chi_-) (\delta\chi_+ - \delta\chi)  [\delta\chi_3 - (1-q)\delta\chi] \,,
\end{align}
where $\delta\chi_-$, $\delta\chi_+$, and $\delta\chi_3/(1-q)$ are the roots of $\Sigma^2 (\delta\chi)$. The prefactor $\mathcal{A}=\sqrt{\bar\sigma \sigma_3/(1-q)}$ is given by
\begin{align}
\mathcal{A} &= \frac{3}{2} \frac{1}{\sqrt{(1+q)}} \left(\frac{r}{M}\right)^{-11/4} \left( 1 - \frac{\chi_{\rm eff}}{\sqrt{r/M}}\right)\geq0\,.
\end{align}
 The conditions $\bar \sigma\geq 0$ and $ \sigma_3\geq0$ imply that the only bounded region where $\Sigma^2 (\delta\chi)\geq 0$ lies between the two smaller roots $\delta\chi_-$ and $\delta\chi_+$ while the spurious solution $\delta\chi_3/(1-q)$ must necessarily be the largest of the three (see Fig.~\ref{poly}). Physical values of $\delta\chi$ describing spin precession must satisfy
 \begin{align}
 \label{rootordering}
\delta\chi_- \leq \delta\chi \leq \delta\chi_+ \leq \frac{\delta\chi_3}{1-q}\,.
\end{align}
Crucially, in this formulation the quantities $\delta\chi_{-,+,3}$ do not have hidden divergences and remain finite when $q\to 1$ (see Sec.~\ref{notable}).
For some of the following expressions, it is useful to perform an affine transformation and identify a binary with the parameter 
 \begin{align}
 \label{deltachitilde}
{\delta\tilde\chi} = \frac{\delta\chi - \delta\chi_-}{\delta\chi_+ - \delta\chi_-} \in [0,1]\,.
 \end{align} 

A fast and accurate determination of $\delta\chi_{-}$, $\delta\chi_+$ and $\delta\chi_3$ from the coefficients $\sigma_i$ is crucial for a successful numerical implementation. While the solution of a third-degree polynomial is analytical, the resulting algebraic expressions are convoluted and a standard numerical algorithm based on the eigenvalues of the companion matrix~\cite{horn2012matrix,2020Natur.585..357H} appears to perform better (see Appendix~\ref{rootsalg}). This is a considerable improvement compared to our previous implementation~\cite{2016PhRvD..93l4066G}, where the turning points were determined with a custom root finder based on effective potentials \cite{2015PhRvL.114h1103K}.  The exploitation of the algebraic properties of ${\dd \delta \chi }/{\dd t}$ is a key element of the computational speedup achieved with the current version of our code (Sec.~\ref{pythonsec}). 
More specifically, we solve the cubic polynomial twice. We first solve $\Sigma=0$ in $\delta\chi$ and retain the two smaller roots. We then solve $(1-q)^2 \Sigma=0$ in $\delta\chi' = (1-q) \delta\chi$ and retain the larger root. Because $\sigma_3\propto(1-q)$, this allows us to compute the three quantities $\delta\chi_{-,+,3}$ regularly for all values of $q\leq 1$. %

\begin{figure}
\includegraphics[width=\columnwidth]{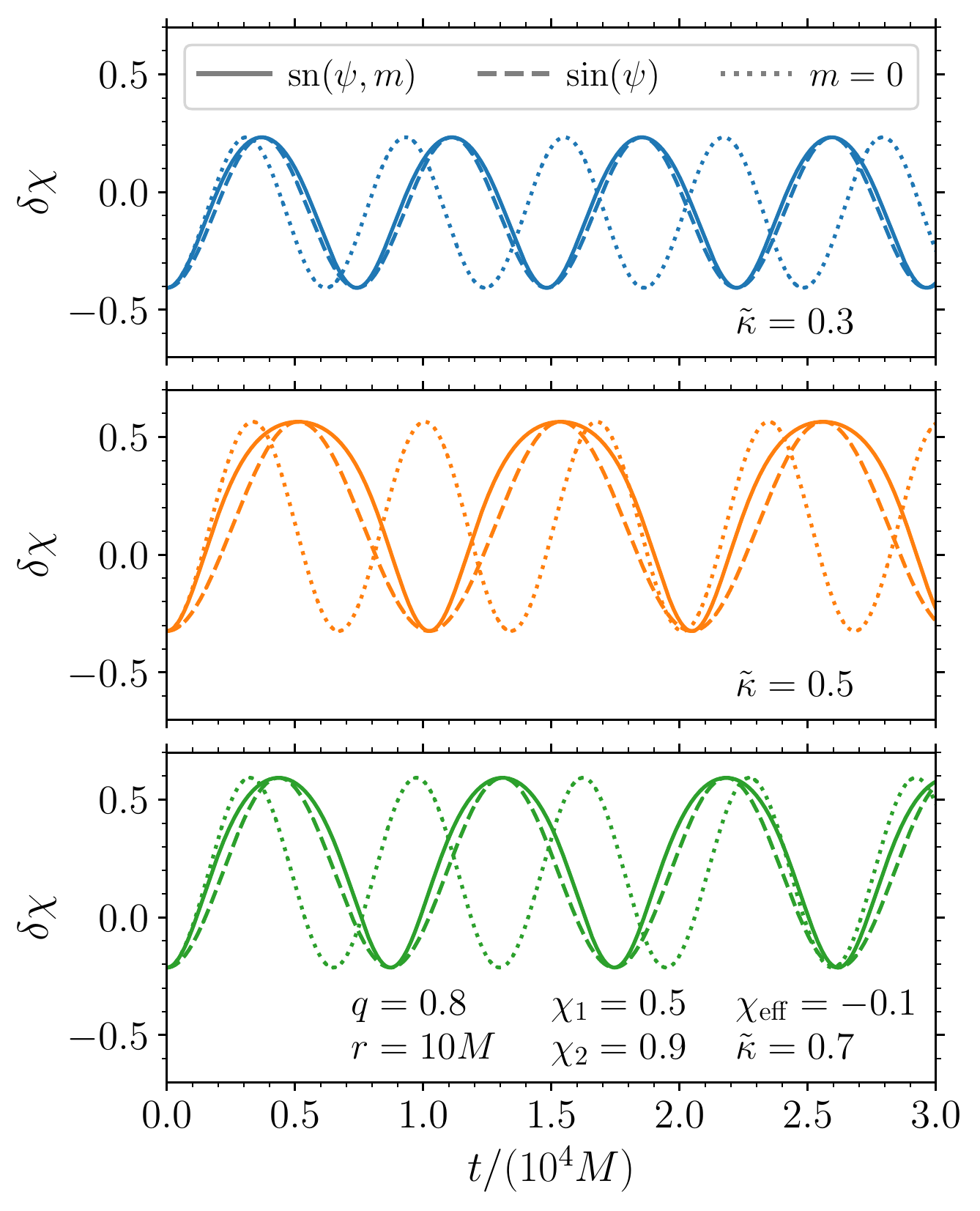}
\caption{Evolution of the weighted spin difference $\delta\chi$ as a function of time on the precession timescale (i.e. neglecting radiation reaction). All three binaries have the same values of $q=0.8$, $\chi_1=0.5$, $\chi_2=0.9$, $\chi_{\rm eff}=-0.1$, and $r=10M$, but three different values of $\tilde\kappa=0.3$ (blue, top panel), $0.5$ (orange, middle panel), and $0.7$ (green, bottom panel). Solid curves show the full solution which is provided in terms of Jacobi elliptic functions. Dotted curves show approximate solutions where we set $m=0$. Dashed curves show approximate solutions where we use standard trigonometric functions but keep the period of the oscillation as derived from the full solution.
}
\label{jacobisolutions}
\end{figure}

We can integrate Eq.~(\ref{cubicandroots}) in time, setting $\delta\chi(t=0)=\delta\chi_-$ as the initial condition. The formal solution is (cf.~\cite{2017PhRvL.118e1101C,2017PhRvD..95j4004C,2021arXiv210610291K})
\begin{align}
\label{jacobisolution}
&\delta\chi =\delta\chi_- +(\delta\chi_+ - \delta\chi_-)  \notag \\
 &\times \sn^2 
 \left[ {\frac{\mathcal{A}}{2} \sqrt{\delta\chi_3-(1-q)\delta\chi_-} \frac{t}{M}},   \frac{(1-q)(\delta\chi_+-\delta\chi_-)}{\delta\chi_3- (1-q) \delta\chi_-} \right] \,,
\end{align}
where $\sn(\psi, m)$ is the Jacobi elliptic sine   \cite{1965hmfw.book.....A}. %
In a nutshell, $\sn(\psi ,m)\in[-1,1]$ is a periodic function with $\psi$-period $4 K(m)$, where $K(m)$ is the complete elliptic integral of the first kind. The elliptic sine is qualitatively similar to the standard trigonometric sine and reduces exactly to it for $m=0$.  In our case, the elliptic parameter is  
\begin{align}
\label{ellipticm}
m &=  \frac{(1-q)  (\delta\chi_+-\delta\chi_-)}{\delta\chi_3- (1-q) \delta\chi_-}\,.
\end{align}
The period of $\sn^2$ is $2 K(m)$, which implies that the weighted spin difference $\delta\chi$  oscillates from $\delta\chi_-$ to  $\delta\chi_+$ and back to $\delta\chi_-$ in a time given by
\begin{align}
\label{tauperiod}
\frac{\tau}{M}  =
\frac{ 4 K(m) }{\mathcal{A} \sqrt{\delta\chi_3-(1-q)\delta\chi_-}}\,.
\end{align}
The variable $\tau$ in Eq.~(\ref{tauperiod}) is the spin nutational period as in Refs.~\cite{2015PhRvL.114h1103K,2015PhRvD..92f4016G}. %

Using the rescaling put forward in Eq.~(\ref{deltachitilde}),
 one can rewrite Eq.~$(\ref{jacobisolution})$ as  
\begin{align}
\label{jacobisolution2}
\delta\tilde\chi = \sn^2 \left[ 2 K(m)  \frac{ t }{\tau} ,m \right]\,.
\end{align}
This expression can then be inverted to obtain
\begin{align}
\label{tofdeltachi}
t =
\pm \,\frac{\tau}{2} \, \frac{F\left(\arcsin\sqrt{\delta\tilde\chi}, m\right)}{K(m)}
\,,
\end{align}
where $F(\varphi,m)$ is the incomplete elliptic integral of the first kind~\cite{1965hmfw.book.....A},
whose definition implies that $F(\pi/2,m) = K(m)$. 
The $\pm$ sign in front refers to the two halves of the cycle, $\delta\chi_-\to \delta\chi_+$ and $\delta\chi_+\to\delta\chi_-$. One has $t(\delta\chi_-)=0$ and $t(\delta\chi_+)=\pm \tau/2$, such that a full oscillation takes a time $\tau$.
Equation~(\ref{tofdeltachi}) can also be derived directly from Eq.~(\ref{cubicandroots})  using some of the standard integrals reported in Appendix~\ref{appintegrals}. When tackled this way, the oscillation period is most naturally given by the integral  $\tau = 2 \int_{\delta\chi_-}^{\delta\chi_+} (\dd \delta\chi/\dd t)^{-1} \dd\delta\chi$, which is the calculation that was performed numerically in our previous implementation~\cite{2016PhRvD..93l4066G}.

Some of the resulting solutions are shown in Fig.~\ref{jacobisolutions} for a set of three binaries that share the same values of the constants of motion. In particular, we show the full solution of Eqs.~(\ref{jacobisolution}) and (\ref{jacobisolution2})  together with two possible approximations (see e.g. Ref.~\cite{2022arXiv221204657L}). Setting $m=0$ in both Eqs.~(\ref{tauperiod}) and (\ref{jacobisolution2}) results in very large deviations, with the spin going several radians out of phase in just a few cycles. One can instead set $m=0$ only in Eq.~(\ref{jacobisolution2}) and not in Eq.~(\ref{tauperiod}), i.e., approximate $\sn(\psi,m)\simeq \sin(\psi)$ for the shape of the function but ensure that the period $\tau$ is the same. For the binaries considered in  Fig.~\ref{jacobisolutions}, we find this procedure results in errors on $\delta\chi$ that are $\lesssim 0.2$. We anticipate this second approximation could be useful in waveform construction, though one should first explore its accuracy more extensively in the parameter space. %

Precession-averaged evolutions require a final resampling of the precessional phase  (see Sec.~\ref{bininsp}). This task is now straightforward:  instead of relying on inverse-transform sampling as in Ref.~\cite{2016PhRvD..93l4066G}, one can simply generate a random number $t$ uniformly in  $[0,\tau]$ and evaluate Eq.~(\ref{jacobisolution}) to obtain $\delta\chi$.  %

\subsection{Parameter boundaries and spin--orbit resonances}
\label{secbounds}

The various parameters describing BH-binary spin precession are bounded by several constraints. We discuss their limits starting from the constants of motion, then moving on to quantities that vary on the long and short timescales of the problem.%

\subsubsection{Mass ratio} We use a convention where labels ``1'' and ``2'' indicate the heavier and lighter BHs, respectively. The mass ratio $q=m_2/m_1$ is defined in $(0,1]$.

\subsubsection{Spin magnitudes}
The Kerr geometry imposes $\chi_{1,2}\in[0,1]$. More conservatively, the spins of astrophysical BHs are not expected to exceed the \citeauthor{1974ApJ...191..507T}
limit $\chi_{1,2} \lesssim 0.998$~\cite{1974ApJ...191..507T}.

\subsubsection{Effective spin}
From Eq.~(\ref{chieff}), the effective spin $\chi_{\rm eff}$ is strictly defined in $[-1,1]$. Fixing $q,\chi_1$, and $\chi_2$ further restricts the allowed range of $\chi_{\rm eff}$ to
 \begin{align}
 \label{chieffdef}
 - \frac{\chi_1 + q \chi_2}{1+q} \leq \chi_{\rm eff} \leq   \frac{\chi_1 + q \chi_2}{1+q}\,.
\end{align}

\subsubsection{Orbital separation} 
A conservative threshold for the PN approximation to be valid is $r>10 M$, or equivalently $u \leq (1+q)^2 / (2q  \sqrt{10})$~\cite{2006PhRvD..74j4005B,2009PhRvD..80h4043B,2009PhRvD..79h4010C}. While astrophysical binaries form at some large but finite separation~\cite{2022PhRvD.106b3001J}, considering the asymptotic behavior at $r\to \infty$ has key applications in GW population studies~\cite{2022PhRvD.105b4076M}.

\subsubsection{Asymptotic angular momentum} From Eq.~(\ref{kappadef}), the geometrical constraint $\boldsymbol{J}=\boldsymbol{L}+\boldsymbol{S}_1+\boldsymbol{S}_2$ translates into
 \begin{align}
 \label{kappabounds}
 \begin{dcases}\frac{  \kappa}{M^2}&\geq   \frac{q \sqrt{r/M}}{2 (1+q)^2}  \Bigg\{ \max  \Bigg[0,\\
  & \left(1- \frac{\chi_1 + q^2 \chi_2}{q \sqrt{r/M}} \right) \left|1- \frac{\chi_1 + q^2 \chi_2}{q \sqrt{r/M}} \right|,\\   &   \left(\frac{|\chi_1 - q^2 \chi_2|}{q \sqrt{r/M}} -1  \right) \left|\frac{|\chi_1 - q^2 \chi_2|}{ q \sqrt{r/M}} -1  \right|  \Bigg]-1 \Bigg\}\,, \\
 \frac{\kappa}{M^2}&\leq \frac{\chi_1 + q^2 \chi_2}{(1 + q)^2} \left(\frac{\chi_1 + q^2 \chi_2}{2 q \sqrt{r/M}} + 1\right)\,,\end{dcases} \end{align}    
where the first condition corresponds to $J=0$,
 the second condition corresponds to 
 $\cos\theta_1=\cos\theta_2=-1$ (down--down),
 the third condition corresponds to 
  $\cos\theta_1=-\cos\theta_2=\pm 1$ (either up--down or down--up),
  and the fourth condition corresponds to 
$\cos\theta_1=\cos\theta_2=1$ (up--up).
 
The interval reported in Eq.~(\ref{kappabounds}) corresponds to the bounds on $\kappa$ for given values of $q, \chi_1, \chi_2,$ and $r$. However, this range is not available to each binary in its entirety, as that depends on the additional constant of motion $\chi_{\rm eff}$. Section~\ref{rootssubsec} illustrated that spin precession can only take place when the cubic $\Sigma(\delta\chi)$ admits three roots, i.e., when its discriminant is positive. More precisely, we indicate the discriminant of $\Sigma/\bar\sigma$ with $\Delta$. This turns out to be a fifth-degree polynomial in $\kappa$ \cite{2020PhRvD.101l4037M}:
\begin{align}
\label{deltakappa}
\Delta(\kappa) &= 
{\sigma_2^{2}\sigma_1^{2}-4\sigma_3\sigma_1^{3}-4\sigma_2^{3}\sigma_0-27\sigma_3^{2}\sigma_0^{2}+18\sigma_3\sigma_2\sigma_1\sigma_0}
 \\
&= \bar\delta \bigg[
\delta_5  \left(\frac{\kappa}{M^2}\right)^5 +
\delta_4  \left(\frac{\kappa}{M^2}\right)^4+
\delta_3  \left(\frac{\kappa}{M^2}\right)^3
\notag \\
&
\qquad+\delta_2  \left(\frac{\kappa}{M^2}\right)^2+
\delta_1  \left(\frac{\kappa}{M^2}\right)+
\delta_0
\bigg]\,,
\end{align}
where the coefficients $\bar\delta$ and $\delta_i$ are lengthy but algebraic expression involving $q,\chi_1,\chi_2,\chi_{\rm eff}$, and $r$; see Appendix~\ref{polycoeffs}. For convenience, we collect a positive term $\bar\delta$ and isolate the leading-order coefficient $\delta_5=-uM^2$. 
A few examples of $\Delta$ as a function of $\kappa$ are shown in the right panel of Fig.~\ref{poly}.

The roots of the equation $\Delta(\kappa)=0$ correspond to locations in the parameter space where $\delta\chi_- = \delta\chi_+$. Physically, these are cases where the relative orientation of the spins and the orbital angular momentum is fixed on the precession timescale. These configurations are the so-called ``spin--orbit resonances'' first discovered by \citeauthor{2004PhRvD..70l4020S}~\cite{2004PhRvD..70l4020S} and later explored at length by several authors \cite{
2010PhRvD..81h4054K,
2010ApJ...715.1006K,
2012PhRvD..85l4049B,
2013PhRvD..87j4028G,
2014PhRvD..89l4025G,
2014CQGra..31j5017G,
2015PhRvL.114h1103K,
2015PhRvD..92f4016G,
2016MNRAS.457L..49C,
2016PhRvD..93d4071T,
2018PhRvD..98h3014A,
2020PhRvD.101l4037M,
2022PhRvL.128c1101V}. In particular,  Ref.~\cite{2020PhRvD.101l4037M} formally proved that there are always two spin--orbit resonances $\kappa_{\pm}$ for each set of $(q,\chi_1,\chi_2, r,\chi_{\rm eff})$, as previously suggested by extensive numerical explorations~\cite{2004PhRvD..70l4020S,2015PhRvD..92f4016G}. 

Spin precession takes place in a compact interval 
\begin{align}
\label{mkappap}
\kappa_-\leq \kappa \leq \kappa_+\,,
\end{align}
where $\Delta(\kappa)\geq 0$ (see Fig.~\ref{poly}). The quintic equation $\Delta(\kappa)=0$ admits either one, three, or five real roots. We can discard the case with a single real root because there must always be two resonances~\cite{2020PhRvD.101l4037M}. Because $\delta_5<0$, if there are three real roots, the only bounded and positive interval is located between the two greater roots. These are indeed the resonances $\kappa_\pm$ while the third, smaller root is spurious. The occurrence of five real roots instead provides two bound intervals where $\Delta(\kappa)\geq 0$; if the roots are ordered in $\kappa$, there are two pairs of candidate resonances, namely the second and third roots as well as the fourth and fifth roots. Knowing that only one of such pairs can correspond to $\kappa_\pm$ \cite{2020PhRvD.101l4037M}, we calculate the corresponding ranges in $\delta\chi$ and select among them by imposing the constraint of Eq.~(\ref{rectangle}).
 
To compare binaries with different values of the constant of motions (see e.g. Fig.~\ref{Svsdeltachifig} where we vary $q$), it is useful to define
\begin{align}
\label{kappatilde}
\tilde\kappa = \frac{ \kappa - \kappa_-}{\kappa_+ - \kappa_-}\in[0,1]\,.
\end{align} 

With an analogous calculation to that we just presented, one can also expand $\Delta$ as a function of $\chi_{\rm eff}$ to obtain the limits of $\chi_{\rm eff}$ constrained to $q, \chi_1, \chi_2, r$, and $\kappa$. This is less relevant because $\chi_{\rm eff}$ is a constant of motion and should be imposed before $\kappa$, not vice versa.

\subsubsection{Weighted spin difference}
\begin{figure*}
\includegraphics[width=\textwidth]{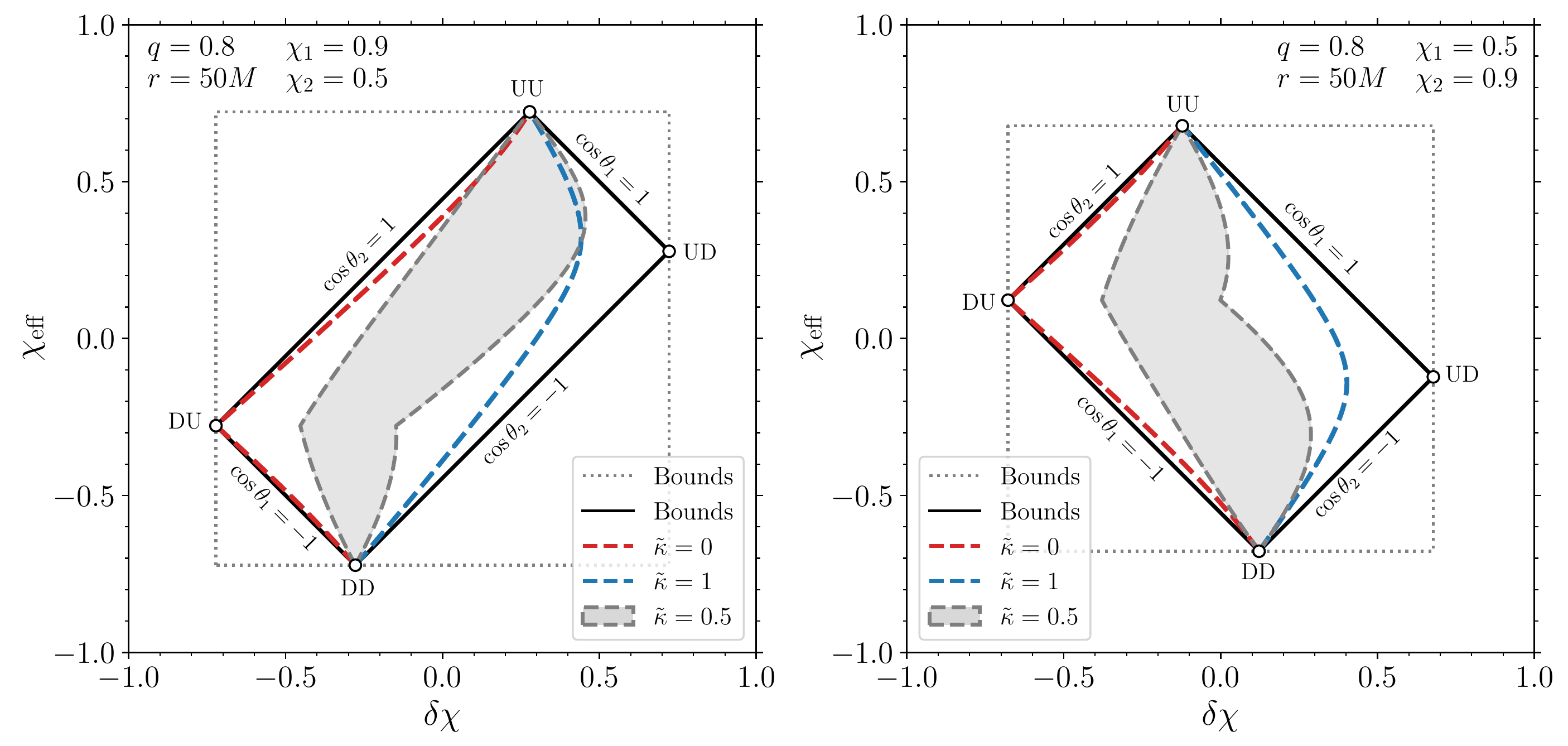}
\caption{Available region in the $\delta\chi-\chi_{\rm eff}$ parameter space for two sets of binaries with fixed values of $q$, $\chi_1$, $\chi_2$, and $r$. The black rectangles indicate the bounds from Eq.~(\ref{rectangle}). Each edge corresponds to one of the four aligned conditions, $\cos\theta_{1,2}=\pm 1$. The rectangles themselves are inscribed in wider squares (gray dotted lines) corresponding to Eqs.~(\ref{chieffdef}) and (\ref{deltachidef}). The parameters used in the left (right) panel have been chosen such that $\chi_1-q\chi_2>0$ ($\chi_1-q\chi_2<0$), which sets the orientation of the rectangle. The white circles at the vertices correspond to the four aligned configurations up--up, up--down, down--up, and down--down (where ``U'' stands for up and ``D'' stands for down). The dashed red and blue curves indicate binaries with $\tilde\kappa=0$ and $\tilde\kappa=1$, respectively, which are the spin--orbit resonances. The gray areas indicate the parameter space available to binaries with $\tilde\kappa=0.5$. For these same binaries, the gray dashed curves mark the turning points $\delta\chi_-$ (left edge) and $\delta\chi_+$ (right edge). }
\label{rectanglefig}
\end{figure*}

From Eq.~(\ref{deltachi}), one has $\delta\chi\in[-1,1]$. Fixing $q,\chi_1,\chi_2$ imposes
\begin{align}
\label{deltachidef}
 - \frac{\chi_1 + q \chi_2}{1+q} \leq  \delta\chi \leq   \frac{\chi_1 + q \chi_2}{1+q}\,.
\end{align}
For a given value of $\chi_{\rm eff}$, the  geometrical conditions $\cos\theta_{1,2} \in [-1,1]$ further restricts the range of $\delta\chi$ to 
\begin{align}
\begin{dcases}
\delta\chi \geq
\max\left( -\chi_{\rm eff} - \frac{2 \chi_1}{1+q} , \chi_{\rm eff} - \frac{2 q \chi_2}{1+q}  \right) \,,
\\
\delta\chi\leq \min\left (  -\chi_{\rm eff} + \frac{2 \chi_1}{1+q}, \chi_{\rm eff} + \frac{2 q \chi_2}{1+q} \right) \,.
\end{dcases}
\label{rectangle}
\end{align}
The resulting constraints are illustrated in Fig.~\ref{rectanglefig} for two representative cases. The allowed region in the $\delta\chi-\chi_{\rm eff}$ plane is a rectangle whose orientation depends on the sign of $\chi_1-q\chi_2$. 

If one also fixes $\kappa$ in addition to $q$, $\chi_1$, $\chi_2$, and $\chi_{\rm eff}$, the evolution of $\delta\chi$ can be further confined to the interval 
\begin{align}
\label{mdchip}
\delta\chi_-\leq \delta\chi \leq \delta\chi_+
\,,
\end{align}
or equivalently $\tilde\delta\chi\in [0,1]$, as discussed in Sec.~\ref{rootssubsec}. These intervals are shown in Figs.~\ref{Svsdeltachifig} and \ref{poly} for a representative set of binaries.  Equation~(\ref{mdchip}) is more restrictive than Eq.~(\ref{rectangle}), which implies that physical regions when spin precession occurs must lie inside one of the rectangles of Fig.~\ref{rectanglefig} but do not occupy them fully.  In particular, Fig.~\ref{rectanglefig} shows the region in the $\delta\chi-\chi_{\rm eff}$ parameter space that is available to binaries with different values of $\tilde\kappa$. Binaries with $\tilde\kappa=0$ or $\tilde\kappa=1$ (i.e., the spin--orbit resonances) are confined to one-dimensional curves because $\delta\chi_-=\delta\chi_+$. On the other hand, binaries with generic values of  $\tilde\kappa\in(0,1)$ occupy a wider, non-degenerate portion of the parameter space.

\subsection{Dynamics in an inertial frame}

While most astrophysical applications only require the mutual orientations of the spins and the orbital angular momentum, tracking the dynamics in an inertial frame is crucial to construct waveforms for GW data analysis. 

The direction of $\boldsymbol{J}$ is constant on the precession timescale, which implies one can use this vector to define an inertial frame and describe the dynamics accordingly. In this frame, the direction of the orbital angular momentum $\boldsymbol{L}$ is defined by a polar angle given by
\begin{align}
\cos\theta_L = \hat{\boldsymbol{L}}\cdot \hat{\boldsymbol{J}}
\end{align}
and an azimuthal angle 
\begin{align}
\Phi_L = \int\Omega_L \dd t
\end{align}
which is measured in the plane orthogonal to $\boldsymbol{J}$. The latter can be found by integrating the precession frequency 
\begin{align}
\Omega_L =\frac{\dd \hat{\boldsymbol{L}}}{\dd t}  \cdot \frac{\hat{\boldsymbol{J}} \times \hat{\boldsymbol{L}}\;\,}{\left|\hat{\boldsymbol{J}} \times \hat{\boldsymbol{L}} \right|^2}
\end{align} 
while neglecting GW emission (i.e., setting $\dd L/\dd t = 0$).

All these quantities can be expressed using the parametrization adopted in Sec.~\ref{lookfor}. One has \cite{2015PhRvL.114h1103K,2015PhRvD..92f4016G,2017PhRvD..96b4007Z,2021PhRvD.103l4026G}
\begin{align}
\label{thetaL}
\cos\theta_L  &= \left[1 + (1 + q) \frac{ (1 - q) \delta\chi + (1 + q) \chi_{\rm eff}}{
    2 q \sqrt{r/M}} \right] 
   \notag \\
   & \times \left[1+\frac{2 (1+q)^2  }{q} \frac{\kappa/M^2}{\sqrt{r/M}}\right]^{-1/2}
\end{align}
and
\begin{align}
\label{thisomegaL}
\Omega_L M= \mathcal{C}_0 \left(1- \sum_{i=\{+,-\}} \frac{\mathcal{C}_i}{\mathcal{R}_i -  (1-q) \delta\chi\sqrt{M/r}}  \right)\,,
\end{align}
where the following coefficients do not depend on $\delta\chi$ but only on quantities that are constant on the precession timescale:\\
\begin{align}
\mathcal{C}_0 &= \frac{q}{2(1+q)^2} \left(\frac{M}{r}\right)^{5/2} 
\sqrt{1+\frac{2 (1+q)^2  }{q} \frac{\kappa/M^2}{\sqrt{r/M}}}\,,
\\
\mathcal{C}_\pm &= \pm 3 \left(1 - \frac{\chi_{\rm eff}}{r/M}\right) 
 \Bigg[ (1 + q)  \left(1 + \frac{\chi_{\rm eff}}{r/M}\right)
 \notag \\&
 \times \!
 \left(1 \pm\sqrt{1\!+\!\frac{2 (1\!+\!q)^2  }{q} \frac{\kappa/M^2}{\sqrt{r/M}}}\right)
\!+ \! \frac{(1\! +\! q)^3}{q}  \frac{\kappa/M^2}{\sqrt{r/M}} 
\notag \\&
-  \frac{1 - q}{2  q^2}\, \frac{\chi_1^2 - q^4 \chi_2^2}{r} \Bigg] 
\left(1+\frac{2 (1+q)^2  }{q} \frac{\kappa/M^2}{\sqrt{r/M}} \right)^{\!\!-1/2}\,,
\\
\mathcal{R}_\pm &= -\frac{2 q}{1+q}  \left(1 \pm\sqrt{1+\frac{2 (1\!+\!q)^2  }{q} \frac{\kappa/M^2}{\sqrt{r/M}}}\right) 
\notag \\
&- (1+q) \frac{\chi_{\rm eff}}{\sqrt{r/M}}\,.
\label{bigR}
\end{align}
One can then integrate
Eq.~(\ref{thisomegaL})
to obtain
\begin{align}
&\Phi_L = %
\pm \mathcal{C}_0 \frac{\tau}{2 M K(m)} 
\Bigg[ 
F\left(\arcsin\sqrt{\delta\tilde \chi}, m \right)
\notag \\ &
-\!\!\!\!
\sum_{i=\{+,-\}}
\!\!\!
 \frac{\mathcal{C}_{} \; n_i}{
 \delta\chi_+\!-\delta\chi_- }  \Pi\left(
n_i\,(1\!-\!q)\sqrt{M/r}  ,
 \arcsin\sqrt{\delta\tilde \chi}, m \right)\!
 \Bigg],
 \label{phil}
\end{align}
where
\begin{align}
n_i =  
 \frac{ \delta\chi_+-\delta\chi_-}{\mathcal{R}_i  -\delta\chi_- (1-q)\sqrt{M/r}}
\,,
\end{align}
and $\Pi(n,\varphi,m)$ is the incomplete elliptic integral of the third kind~\cite{1965hmfw.book.....A}; cf. Appendix~\ref{appintegrals}. The total angle spanned during a nutation period $\tau$ is given by
\begin{align}
\alpha =  \mathcal{C}_0 \frac{\tau}{M}\!
\left\{
\!1
-\!\!\!\!\!\sum_{i=\{+,-\}}\!\!
 \frac{\mathcal{C}_i \; n_i}{
 \delta\chi_+\!-\!\delta\chi_- }  \frac{\Pi\left[n_i 
(1\!-\!q)\sqrt{M/r} \,  m \right]}{K(m)}
 \right\}\!,
\end{align}
where $\Pi(n,m)$ is the complete elliptic integral of the third kind.
Much like in Eq.~(\ref{tofdeltachi}), we have assumed an initial condition such that $\Phi_L(\delta\chi_-) = 0$ and $\Phi_L(\delta\chi_+) = \pm \alpha/2$, where the $\pm$ sign refers to the two halves of the nutation cycle. 

The direction of $\boldsymbol{J}$ is approximately constant also on the longer radiation-reaction timescale (see, e.g.,~\cite{2009ApJ...704L..40B,2010PhRvD..81h4054K,2016ApJ...825L..19H}). Exceptions include the so-called ``nutational resonances''~\cite{2017PhRvD..96b4007Z}, where $\alpha=2\pi n$ for integer $n$. It turns out that  for $n>1$ the resulting tilts in $\boldsymbol{J}$ are as small as $\mathcal{O}(10^{-3})$ rad. The  $n=0$ nutational resonance corresponds to the ``transitional precession'' phenomenon \cite{1994PhRvD..49.6274A} where $J\sim 0$ and tilts are of  $\mathcal{O}(1)$ rad.

\section{Precession-averaged inspiral}
\label{precavinspiral}

\subsection{Precession averaging}
\label{precavrules}

The solutions of Sec.~\ref{rootssubsec} allow us to define the
``{precession average}'' operation.
The precession average of a generic quantity $X$ is given by
\begin{align}
\label{Xbrackets}
\langle X  \rangle = \frac{\displaystyle \int_{\delta\chi_-}^{\delta\chi_+} X (\delta\chi)\left(\frac{\dd \delta\chi}{\dd t}\right)^{-1} \dd \delta\chi }
{\displaystyle \int_{\delta\chi_-}^{\delta\chi_+}  \left(\frac{\dd \delta\chi}{\dd t}\right)^{-1} \dd \delta\chi }
\,,
\end{align}
where the denominator is also equal to $\tau/2$.
In words, we weight each contribution entering $X$ using the ``speed'' $\dd\delta\chi /\dd t$. Imagine taking snapshots of the dynamics over a period $\tau$, one is more (less) likely to measure a given value $X(\delta\chi)$ if the variation of $\delta\chi$ is slow (fast). Note that we can safely integrate only over the first half of a precession cycle $\delta\chi_-\to \delta\chi_+$ because the second half $\delta\chi_+\to \delta\chi_-$ is identical up to a sign change of the derivative $\dd\delta\chi/\dd t$. If the quantity $X$ is constant on the precession timescale (i.e., $\dd X/\dd \delta\chi=0$), one obviously has $\langle X  \rangle=X$. 

\begin{figure}
\includegraphics[width=\columnwidth]{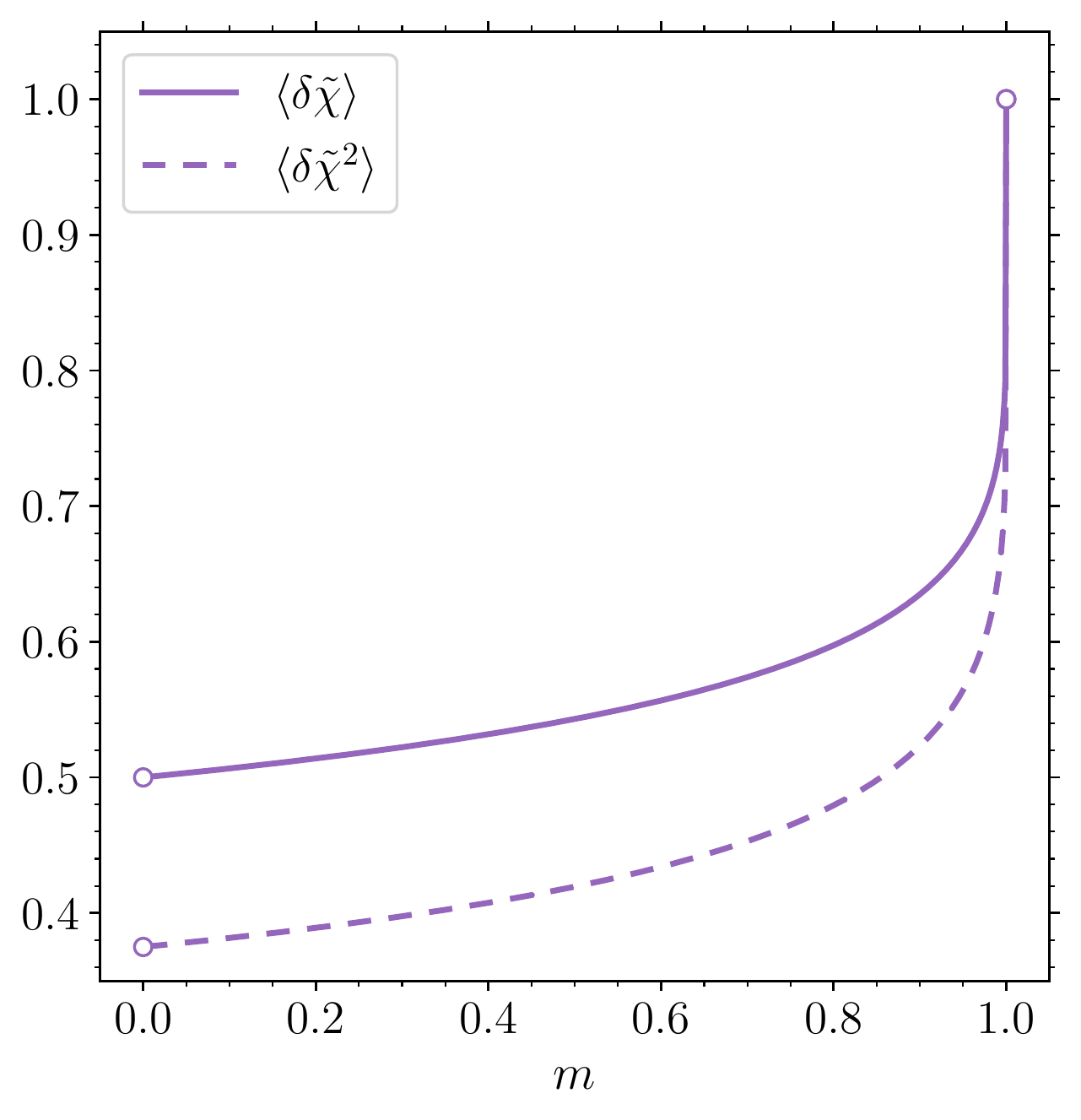}
\caption{Moments of the weighted spin difference $\delta\tilde\chi$ rescaled using the turning points and averaged over a precession cycle. %
Solid and dashed lines show $\langle\delta\tilde\chi\rangle$  and $\langle\delta\tilde\chi^2\rangle$, respectively, as a function of the elliptic parameter $m$; cf. Eqs.~(\ref{deltachitildeav}) and (\ref{deltachitildeav2}).
}
\label{mfunction}
\end{figure}

Using this notion of average, the first two moments of $\delta\tilde\chi$ can be elegantly reduced to special functions; see 
Appendix~\ref{appintegrals}. We find 
\begin{align}
\label{deltachitildeav}
\langle \delta\tilde\chi \rangle &= \frac{1}{m}\left[ 1- \frac{E(m)}{K(m)} \right]\,,
\\
\label{deltachitildeav2}
{\langle \delta\tilde\chi ^2\rangle} &= \frac{1}{3 m^2} \left[2 + m - 2  (1 + m) \frac{E(m)}{K(m)}\right]\,,
\end{align}
where $m$ is given by Eq.~(\ref{ellipticm}) and $K(m)$ and $E(m)$ are the complete elliptic integrals of the first and second kind, respectively (cf. analogous expressions in Ref.~\cite{2021arXiv210610291K}). These quantities are shown in Fig.~\ref{mfunction}. As $m$ increases from 0 to 1, both moments increase monotonically from $\lim_{m\to 0 }\langle \delta\tilde\chi \rangle =1/2$ and $\lim_{m\to 0}\langle \delta\tilde\chi^2 \rangle =3/8$ to $\lim_{m\to 1}\langle \delta\tilde\chi \rangle  = \lim_{m\to 1}\langle \delta\tilde\chi^2 \rangle =1$.

\begin{figure*}
\includegraphics[width=\textwidth]{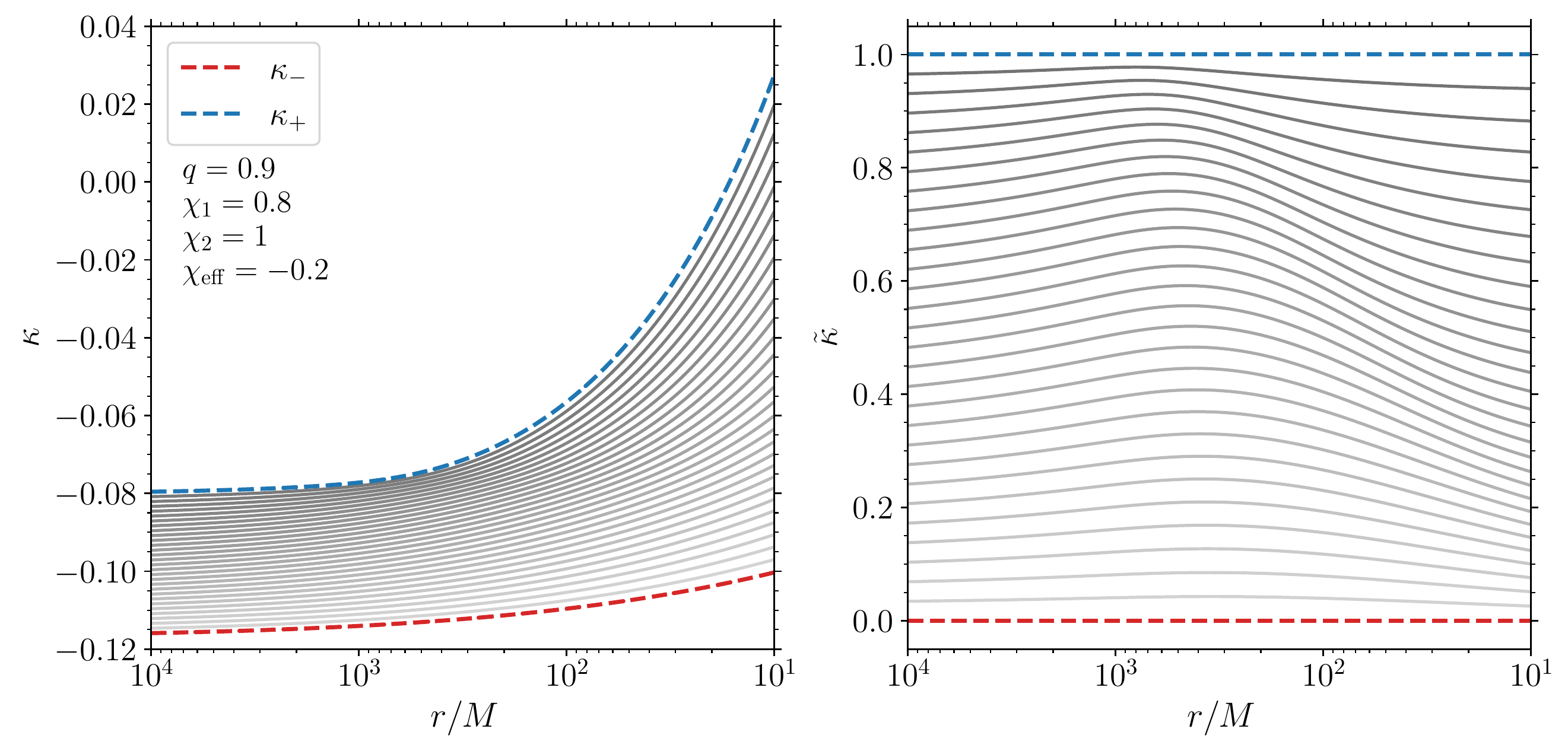}
\caption{Representative integrations of the inspiral dynamics using a precession-averaged approach. The left panel shows the evolution of the asymptotic angular momentum $\kappa$. The right panel shows the same results rescaled to $[0,1]$ using $\tilde\kappa$; see Eq.~(\ref{kappatilde}). The gray curves represent binaries with $q=0.9$, $\chi_1=0.8$, $\chi_2=1$, $\chi_{\rm eff}=-0.2$, and a set of equally spaced initial values of $\tilde\kappa$. These are evolved from $r=10^4M$ to $r=10M$. The spin--orbit resonances $\kappa_{\pm}$ are shown with red and blue dashed curves, respectively.}
\label{kappaandkappatilde}
\end{figure*}

\subsection{Binary inspiral}
\label{bininsp}

The notion of precession average allows us to connect different quantities that vary on the radiation-reaction timescale. The parametrization of Sec.~\ref{lookfor} reveals that, at least for quasi-circular binaries, there are only two such variables: the asymptotic angular momentum $\kappa$ and the orbital separation $r$ (or equivalently the compactified coordinate $u$). The only ingredient one needs to evolve binaries along the inspiral is an ordinary differential equation (ODE) for $\dd\kappa/\dd u$.

As first shown in Refs.~\cite{2015PhRvD..92f4016G,2015PhRvL.114h1103K}, the derivation is straightforward when 
restricting to the angular-momentum flux at 1PN \cite{1995PhRvD..52..821K} and yields
\begin{align}
\label{dkappadu}
\frac{\dd \kappa}{\dd u} = \langle S^2
\,\rangle\,.
\end{align}
The right-hand side can be evaluated using Eqs.~(\ref{Svsdeltachi}) and (\ref{deltachitilde}) and by considering that $\delta\tilde\chi$ is the only variable that evolves on the precession timescale:
\begin{align}
\langle S^2 \rangle &=    \frac{M^2}{2 u} \bigg\{2 \frac{\kappa}{M^2} - \chi_{\rm eff} 
\notag \\
&-  \frac{1-q}{1+q}  \big[ \delta\chi_-  + \langle \delta\tilde\chi \rangle  (\delta\chi_+ -  \delta\chi_-)\big]   \bigg\}
\,.
\label{Svsdeltachiav}
\end{align}
The average $\langle\delta\tilde\chi\rangle$ is given by Eq.~(\ref{deltachitildeav}) and it depends on $m$ from Eq.~(\ref{ellipticm}).

In summary, a precession-averaged binary inspiral is determined by a single ODE of the form $\dd\kappa/\dd u = {\rm RHS}(\kappa,u)$ that needs to be evolved from some initial condition $(u_0,\kappa_0)$. The quantities $\delta\chi_{-,+,3}$ (and hence $m$) entering the right-hand side all depend on both $\kappa$ and $u$ in a non-trivial fashion. While we were not able to find an analytic solution, numerically integrating this Cauchy problem does not present a significant computational challenge. Some examples are shown in Fig.~\ref{kappaandkappatilde}. The solutions $\kappa(r)$ are smooth and lie between the spin--orbit resonances $\kappa_{\pm}$. A less trivial behavior appears when considering the evolution of $\tilde\kappa(r)$, with binaries approaching and departing from the resonances $\tilde\kappa=0,1$ as they inspiral toward merger. 

Previous literature, including by some of us (e.g. Ref.~\cite{2013PhRvD..87j4028G}), have referred to the spin--orbit resonances as attractors because of their impact on the angle $\Delta\Phi$ (see Sec.~\ref{morphologies}). We now believe that this attribute is, to some extent at least, misplaced. When pictured in terms of parameters that respect the timescale separation of the dynamics, BH binaries do not generically approach the spin--orbit resonances as they inspiral toward merger.

The workflow of a precession-averaged inspiral is the following:
\begin{itemize}
\item Assume one is provided an initial binary configuration in terms of the mass ratio $q$ and the initial conditions of the three momenta $\boldsymbol{L}_i$, $\boldsymbol{S}_{1i}$, $\boldsymbol{S}_{2i}$. 
\item From the magnitudes of the momenta, compute $\chi_1$ and $\chi_2$ (which are constant of motions; we thus drop the initial condition subscript $i$) and $r_i$ (or equivalently  $u_i$) from Eqs.~(\ref{spinmags})-(\ref{udef}). 
\item Compute the angles $\theta_{1i},\theta_{2i}$, and $\Delta\Phi_{i}$ from Eqs.~(\ref{costheta1})-(\ref{dphisign}).
\item Convert these three angles into $\chi_{\rm eff}$, $\kappa_i$, and $\delta\chi_i$ from Eqs.~(\ref{chieff})-(\ref{deltachi}). If the binary is provided at infinitely large separation, only $\theta_{1,2}$ (and not $\Delta\Phi$) enter this conversion. 
\item Retain only $\chi_{\rm eff}$ (which is a constant of motion) and $\kappa_i$ (which provides the initial condition for the ODE integration); $\delta\chi_i$ is not necessary. 
\item Integrate $\dd\kappa/\dd u$ from Eq.~(\ref{dkappadu}) to the desired final separation $r_f$, resulting in $\kappa_f$.
\item At the final separation, extract a random value of $t\in [0,\tau]$ and evaluate $\delta\chi_f$ from Eq.~(\ref{jacobisolution}). Also extract a random sign $\varepsilon =\{ -1,+1\}$.
\item Convert $q,\chi_1,\chi_2,\chi_{\rm eff}, r_f, \kappa_f,$ and $\delta\chi_f$ into the angles $\theta_{1f},\theta_{2f},$ and $\Delta\Phi_{f}$ using Eqs.~(\ref{ct1})-(\ref{cosdeltaphi}). While performing this conversion, assume $\Delta\Phi_f = \varepsilon \times \arccos( \cos\Delta\Phi_f)\in[-\pi,\pi]$. This reflects the underlying symmetry of the two halves of a precession cycle. 
\end{itemize}

Note how, in this scheme, one explicitly loses memory of the initial value of $\delta\chi$ and resamples it at the very end. In other words, one does not track the evolution of the spins along their precession cones but only the ``shape'' of those cones along the inspiral. The uncertainty introduced via the precession-averaging procedure can be captured by resampling many final values of $\delta\chi$ and construct distributions of quantities at $r_f$, rather than point estimates  (see, e.g., Ref.~\cite{2020CQGra..37v5005R}).

\section{Phenomenology of spin precession}
\label{phenosec}

\subsection{Some notable limits}
\label{notable}

We now consider some notable limits of the dynamics, namely those at equal masses and large separations.

\subsubsection{Equal masses: $\boldsymbol{q\to 1}$} 

Parametrizing the precession dynamics using $\delta\chi$ allows us to seamlessly study equal-mass systems ---a task that had previously required a separate formulation \cite{2017CQGra..34f4004G}. 

In particular, for $q=1$ one has $\sigma_3=0$ such that the cubic  polynomial $ ({d\delta \chi }/{d t})^2$ from Eq.~(\ref{thirddegree}) reduces to a parabola; see Fig.~\ref{poly}. This is consistent with Eq.~(\ref{rootordering}), with the %
 largest root $\delta\chi_3/(1-q)$ approaching $+\infty$ as $q\to 1$.  
The discriminant $\Delta(\kappa)$ reduces to a cubic polynomial for which the  spin--orbit resonances are the two largest solutions:
\begin{align}
\label{kappaminusqone}
\lim_{q\to 1} \frac{\kappa_-}{M^2}&= \max\left[ \frac{(\chi_1-\chi_2)^2}{8} , \frac{\chi_{\rm eff}^2}{2}\right] \sqrt\frac{M}{r} + \frac{\chi_{\rm eff}}{2}
\,,
\\
\label{kappaplusqone}
\lim_{q\to 1} \frac{\kappa_+}{M^2} &=  \frac{(\chi_1+\chi_2)^2}{8}  \sqrt\frac{M}{r} + \frac{\chi_{\rm eff}}{2}
\,.
\end{align}
 From Eq.~(\ref{ellipticm}), the condition  $q=1$ implies $m=0$. In this case, the time evolution of the BH spins is substantially simpler because $\sn(\psi,0) = \sin(\psi)$ and $K(0)=\pi/2$.
In particular, the nutation period is given by
\begin{align}
\lim_{q\to 1}\tau = \frac{4\pi}{3} \!\left(\frac{r}{M}\right) ^{\!11/4}{\!{\left(2 \frac{\kappa}{M^2} \!-\! \chi_{\rm eff}\right)^{\!-1/2}
}   \! \left( 1 \!-\! \frac{\chi_{\rm eff}}{\sqrt{r/M}}\right)^{\!-1}} \!\!\!\! M
\end{align}
and the total nutation angle is given by
\begin{align}
\lim_{q\to 1} \alpha &= 
\frac{\pi}
{6} \left(\frac{r}{M}\right)^{1/4} 
 \left(7- 6  \frac{\chi_{\rm eff}}{\sqrt{r/M}} \right)
  \left( 1 - \frac{\chi_{\rm eff}}{\sqrt{r/M}}\right)^{-1}
\notag \\
&
\left({1+8\frac{\kappa/M^2}{\sqrt{r/M}}}\right)^{1/2}
{\left(2 \frac{\kappa}{M^2} - \chi_{\rm eff}\right)^{-1/2}
} \,.
\end{align}

Using Eqs.~(\ref{dS1st})-(\ref{OmegaL}), it is immediate to prove \cite{2017CQGra..34f4004G} that $q=1$ implies $\dd S/\dd t = 0$, i.e., the magnitude of the total spin $S$ is a constant of motion on all timescales; as highlighted in Sec.~\ref{lookfor}, this is the reason behind the coordinate singularity that affected our previous formulation~\cite{2015PhRvL.114h1103K,2015PhRvD..92f4016G}. From Eq.~(\ref{Svsdeltachi}), the value of this constant is
\begin{align}
\lim_{q\to1} \frac{S^2}{M^4} =    \frac{1}{4}\sqrt{\frac{r}{M}} \left(2 \frac{\kappa}{M^2} - \chi_{\rm eff} \right)\,.
\label{Svsdeltachiqone}
\end{align}
One has $\langle S \rangle = S$, such that  Eq.~(\ref{dkappadu}) reduces to 
\begin{align}
\lim_{q\to 1} \frac{\dd \kappa}{\dd u} =    \frac{1}{u} \left( \kappa - \frac{\chi_{\rm eff}}{2} M^2  \right)\!.
\end{align}
This can be solved analytically from some initial condition $\kappa(u_0) = \kappa_0$:
\begin{equation}
\lim_{q\to 1} \frac{\kappa(r)}{M^2} = \frac{\chi_{\rm eff}}{2} + \sqrt{\frac{r_0}{r}} \left( \frac{\kappa_0}{M^2} - \frac{\chi_{\rm eff}}{2}  \right)\,,
\label{kapparq1}
\end{equation}
where we have expressed the results in terms of the orbital separation for clarity. The $q=1$ evolutionary flow, therefore, corresponds to the following conservation law
\begin{align}
 \lim_{q\to1} \left(\frac{\kappa}{M^2} - \frac{\chi_{\rm eff}}{2}  \right)\sqrt{\frac{r}{M}} = {\rm const.} 
\end{align}
Equations~(\ref{kappaminusqone}) and (\ref{kappaplusqone}) indicate that the spin--orbit resonances $\kappa_\pm$ obey the same conservation law. This implies that the rescaled quantity $\tilde\kappa$ from Eq.~(\ref{kappatilde}) remains constant on the inspiral timescale. For the case of equal-mass binaries, curves like those shown in the right panel of Fig.~\ref{kappaandkappatilde} would be straight, horizontal lines.

\subsubsection{Large separations: $\boldsymbol{r\to \infty}$} Considering sources at infinitely large separations (i.e., $r\to \infty$ or $u\to 0$) is useful to provide a consistent reference point to label the binary inspiral and combine GW events at the population level~\cite{2022PhRvD.105b4076M,2022PhRvD.106b3001J,2023PhRvX..13a1048A}.
In this limit, spin--spin couplings can be neglected relative to spin--orbit couplings, which implies that spins precess about $\boldsymbol{L}$ tracing cones with fixed opening angles. That is, both angles $\theta_1$ and $\theta_2$ are asymptotically constant as $r\to \infty$, and so is  $\delta\chi$ because of Eq.~(\ref{deltachi}). %

Plugging the definition $\boldsymbol{J}=\boldsymbol{L}+\boldsymbol{S}_1+\boldsymbol{S}_2$ into Eq.~(\ref{kappadef}) yields \cite{2015PhRvD..92f4016G}
\begin{equation}
\label{kappainf} 
\lim_{r\to \infty} \frac{\kappa}{M^2} = \boldsymbol{S}\cdot  \hat{\boldsymbol{L}} = \frac{\chi_1\cos\theta_1 +q^2 \chi_2\cos\theta_2}{(1+q)^2}\,,
\end{equation} 
which implies that the asymptotic angular momentum $\kappa$ is also constant in the large-separation limit and it is equal to the projection of the total spin along the binary orbital angular momentum. %
Indeed, this property is the very reason why we picked $\kappa$ instead of $J$ in Sec.~\ref{lookfor}.  From Eqs.~(\ref{chieff}) and (\ref{kappainf})  one has %
\begin{align}
\lim_{r\to \infty}  \cos\theta_1  &= \frac{(1+q)[\kappa M^{-2} (1+q) - q \chi_{\rm eff}]}{(1-q)\chi_1}\,,
\label{ct1inf}
\\
\lim_{r\to \infty}  \cos\theta_2 &= 
\frac{(1+q)[\chi_{\rm eff} - \kappa M^{-2} (1+q) ]}{(1-q) q \chi_2}\,.
\label{ct2inf}
\end{align}
The large-separation limit is therefore that peculiar location in the parameters space where all three variables $\chi_{\rm eff}$, $\kappa$, and $\delta\chi$ are constant.  In particular, 
combining Eqs.~(\ref{chieff}), (\ref{deltachi}), and (\ref{kappainf}) returns
\begin{align}
\label{dchiinf}
\lim_{r\to \infty} \delta\chi = \frac{1+q}{1-q}\left(2 \frac{\kappa}{M^2} - \chi_{\rm eff}\right)\,.
\end{align}
The latter expression can also be found using Eq.~(\ref{thirddegree}): 
for $r\to \infty$, one has $\sigma_3\to 0$ %
such that $(\dd \delta\chi /\dd t)^2$ becomes a quadratic polynomial with two coincident roots given by Eq.~(\ref{dchiinf}). The scaling with the separation is important here. In particular, the difference between the right- and left-hand sides of Eq.~(\ref{dchiinf}) is equal to the term in parentheses in Eq.~(\ref{Svsdeltachi}) and is related to the magnitude of the total spin $S$. The magnitude $S\leq |S_1+S_2|$  must remain finite at any separation, including $r\to \infty$, which implies
\begin{align}
\label{sconstant}
\lim_{r\to\infty} \left( 2 \frac{\kappa}{M^2} - \chi_{\rm eff} -  \frac{1-q}{1+q}   \delta\chi \right)  = \mathcal{O}\left(\sqrt{ \frac{M}{r}}\right)
\, .
\end{align}
From this expression, one can make sense of the cosines in Eqs.~(\ref{cosdeltaphi}) and (\ref{costheta12}), which indeed do not diverge as $r\to\infty$.

For the same reason,  in the large-separation limit one cannot naively evaluate Eq.~(\ref{Svsdeltachiav}) to compute $\langle S^2\rangle$ and integrate $\dd\kappa/\dd u$. The right-hand side, however, can be computed directly from the geometrical definition of Eq.~(\ref{Sgeodef}). %
At large separations, the two spins move along cones of constant opening angles given by Eqs.~(\ref{ct1inf}) and (\ref{ct2inf}) with different angular velocities (at least in the generic case where $q\neq 1$). The only parameter that varies on the precession timescale is therefore $\Delta\Phi$, which can be used to parametrize the precession cycle (this is not possible at finite values of $r$). One has $\langle \cos\theta_1\rangle$ = const., $\langle \cos\theta_2\rangle$ = const., and $\langle \cos\Delta\Phi\rangle = \int_{-\pi}^\pi \cos\Delta\Phi\, \dd \Delta\Phi / 2\pi= 0$. Plugging these estimates and Eqs.~(\ref{ct1inf}-\ref{ct2inf}) into Eq.~(\ref{Sgeodef}) returns
\begin{align}
\lim_{r\to\infty} \langle S^2 \rangle & = S_1^2+ S_2^2 + 2 S_1 S_2 \cos\theta_1 \cos\theta_2 
 \\
&=  \frac{\chi_1^2 + q^4 \chi_2^2}{(1+q)^4} M^4
-\frac{2 q}{(1-q)^2
   (1+q)^2}
  \notag \\
 &\times   \left[(1+q) \kappa -\chi_{\rm eff} M^2 \right] \left[(1+q) \kappa -q \chi_{\rm eff} M^2 \right]\,.
   \label{Ssavinf}
\end{align}
In our numerical implementation, we rely on this analytic expression whenever the $\dd\kappa/\dd u$ ODE solver attempts a step with $u\leq 0$.

For binaries at infinitely large  separations, the boundaries of the asymptotic angular momentum
for fixed values of $q,\chi_1,$ and $\chi_2$ can be found  by extremizing Eq.~(\ref{kappainf}):
\begin{align}
-  \frac{\chi_1 +q^2 \chi_2}{(1+q)^2} 
\leq \lim_{r\to \infty} \frac{\kappa}{M^2} \leq  \frac{\chi_1 +q^2 \chi_2}{(1+q)^2} \,,
\end{align}
which is indeed the $r\to \infty$ limit of Eq.~(\ref{kappabounds}). The spin--orbit resonances can then be computed by extremizing $\kappa$ under the constrained values of $q,\chi_1, \chi_2,$ and $\chi_{\rm eff}$.  From Eqs.~(\ref{kappainf}) and (\ref{chieff}), one has
\begin{align}
\label{kappainfres}
\lim_{r\to\infty} \frac{\kappa_\pm}{M^2} =
\mathbin{\stackanchor[3pt]{min}{max}}
\bigg[&\frac{q(1+q)\chi_{\rm eff} \pm  (1\!-\!q)\chi_1}{(1+q)^2} ,
\notag \\ 
&\frac{(1+q)\chi_{\rm eff} \pm q(1-q)\chi_2}{(1+q)^2}\bigg]\,,
\end{align}
where $\min$ and $\max$ refer to $+$ and $-$, respectively. Equivalently, taking the $r\to \infty$ limit of the discriminant $\Delta(\kappa)$ returns a quartic polynomial where the four roots are those listed in Eq.~(\ref{kappainfres}); the resonances are then found by excluding the smallest and the largest of the four, which correspond to the min/max operation in front. %

The leading-order expression for the period $\tau$ and the angle $\alpha$ were computed in Refs.~\cite{2017PhRvD..96b4007Z,2021PhRvD.103l4026G} using a geometrical argument similar to the one above. Their result, which we verified with the new formulation, reads
\begin{align}
\lim_{r\to \infty} \tau &=  \frac{4\pi}{3} \left(\frac{r}{M} \right)^{5/2}\frac{1+q}{1-q}\,,
\\
\lim_{r\to \infty} \alpha &= 
\begin{dcases}\frac{2 \pi q(4+3q)}{3(1-q^2)}  \quad &{\rm if}  \qquad \mathcal{Y}\geq 0  \,,
\\
\frac{2 \pi (4q+3)}{3(1-q^2)}  \quad &{\rm if}  \qquad \mathcal{Y}<0\,,
\end{dcases}
\end{align}
where
\begin{align}
\mathcal{Y} = 2 q (1\!+\!q)^3 \frac{\kappa}{M^2}\chi_{\rm eff} - (1\!+\!q)^5\frac{\kappa^2}{M^4} +(1\!-\!q) (\chi_1^2 \!-\!q^4 \chi_2^2)\,.
\end{align}

\subsubsection{Equal masses and large separations} %
 The case of binaries with equal masses at infinitely large separations is delicate. Taking the $r\to\infty$ limit of Eq.~(\ref{Svsdeltachi}) returns $S^2\to \infty$ and taking the $q\to 1$ limit of Eq.~(\ref{dchiinf}) returns $\delta\chi\to \infty$; both results are clearly  unphysical. 

The key point here is that, while both $\kappa$ and $\chi_{\rm eff}$ are constant for $r\to \infty$, they are not necessarily independent of each other. Equation~(\ref{kapparq1})  reveals that 
\begin{align}
\label{qrkappalimit}
\lim_{r\to \infty} \lim_{q \to 1} \frac{\kappa}{M^2} = \frac{\chi_{\rm eff}}{2}\,,
\end{align}
where the order of the limits is important (we consider equal-mass binaries and propagate them back to large separations). 
 The same result can be found from Eqs.~(\ref{chieff}) and (\ref{kappainf}), and Eq.~(\ref{kappainfres}) indicates that the spin--orbit resonances also tend to the same value. The condition of Eq.~(\ref{qrkappalimit}) keeps the $r\to\infty$ limit of  Eq.~(\ref{Svsdeltachiqone}) regular, ensuring that $S$ remains constant. 
 
The unfortunate consequence of Eq.~(\ref{qrkappalimit}) is that the formalism presented in this paper cannot accommodate $q=1$ binaries at $r\to \infty$. For instance, all the $\sigma_i$ from Eq.~(\ref{ddeltachidt}) tend to zero (see Appendix~\ref{polycoeffs}). Consider the Cauchy problem described in Sec.~\ref{bininsp} where, for a given set of constants of motion $(q,\chi_1,\chi_2, \chi_{\rm eff})$, one needs to prescribe an initial condition $\kappa_0$ at $u_0$. If $u_0=0$, there is only one consistent value of $\kappa_0$ as determined by the constant of motion $\chi_{\rm eff}$ from Eq.~(\ref{qrkappalimit}). Physically, binaries can have different spin orientations, but the labeling strategy we use (i.e., $\chi_{\rm eff}$ and $\kappa$) becomes degenerate. This is a similar issue to that addressed (and solved) in Sec.~\ref{lookfor}, where using $\delta\chi$ instead of $S$ cures the $q=1$ coordinate singularity on the precessional timescale. On the radiation-reaction timescale, this coordinate singularity is still present but only at infinitely large separations. Regularizing the joint limits of $q\to 1$ and $r\to\infty$ requires the identification of an inspiral parameter that, unlike $\kappa$, is not uniquely determined by the constants of motion. This investigation is postponed to future work.
 
Let us note that this parameter degeneracy only affects evolutions \emph{from} infinitely large separation. Integrating $\dd \kappa /\dd  u$ with $q=1$ backward \emph{to} past time infinity is a sound operation and simply returns the limit of Eq.~(\ref{qrkappalimit}). However, one cannot then convert the result to $\theta_{1,2}$; cf. Eqs.~(\ref{ct1inf}) and (\ref{ct2inf}), which diverge.

\subsection{$\Delta\Phi$ morphologies}
\label{morphologies}

The binary dynamics on the precession timescale can be classified into the so-called ``spin morphologies,'' according to the behavior of the angle $\Delta\Phi$. These were identified in Refs.~\cite{2015PhRvL.114h1103K,2015PhRvD..92f4016G} and used extensively afterward \cite{2017CQGra..34f4004G,2018PhRvD..98h4036G,2019PhRvD.100l4008P,2019PhRvD..99j3004G,2019CQGra..36j5003G,2020CQGra..37v5005R,2022PhRvD.106f3028S,2022PhRvD.106b4019G,2023arXiv230110125J}.  

Binaries with either $\delta\chi=\delta\chi_-$ or $\delta\chi=\delta\chi_+$ correspond to configurations where the three vectors $\boldsymbol{L}$, $\boldsymbol{S}_1$ and $\boldsymbol{S}_2$ are coplanar. From Eq.~(\ref{cosdphi}), this implies either $\Delta\Phi=0$ or $\Delta\Phi=\pi$. We refer to binaries as ``librating'' (L) if $\Delta\Phi(\delta\chi_-)=\Delta\Phi(\delta\chi_+)$ and ``circulating'' (C) if $\Delta\Phi(\delta\chi_-)\neq \Delta\Phi(\delta\chi_+)$. There are four possible cases:
\begin{itemize}
\item L0: $\Delta\Phi(\delta\chi_-)=\Delta\Phi(\delta\chi_+)=0$;
\item L$\pi$: $\Delta\Phi(\delta\chi_-)=\Delta\Phi(\delta\chi_+)=\pi$;
\item C$+$: $\Delta\Phi(\delta\chi_-)=0$ and $\Delta\Phi(\delta\chi_+)=\pi$;
\item C$-$: $\Delta\Phi(\delta\chi_-)=\pi$ and $\Delta\Phi(\delta\chi_+)=0$.
\end{itemize}
Previous studies on the subject have
grouped
together the C$-$ and C$+$ morphologies into a single C class, though early hints of this distinction can be found in Ref.~\cite{2019CQGra..36j5003G}.

The spin morphology depends on $\kappa, \chi_{\rm eff}, r, q, \chi_1$, and $\chi_2$, but not on $\delta\chi$: it is therefore constant on the precession timescale while radiation reaction can cause transitions between the different classes. We refer the reader to Ref.~\cite{2015PhRvD..92f4016G} for an extensive exploration of these transitions. In a nutshell, morphological transitions take place whenever either $\boldsymbol{S}_1$ or $\boldsymbol{S}_2$ are aligned with $\boldsymbol{L}$ at any point during the precession cycle. Much like the longitude at the Earth's North Pole, the angle $\Delta\Phi$ is instantaneously ill defined if one spin is aligned, allowing for a discontinuous jump between $0$ and $\pi$ at either $\delta\chi_-$ or $\delta\chi_+$. In general, all binaries with $q<1$ belong to the C$+$ class at $r\to\infty$ and tend to transition to the L classes as they inspiral toward merger. Further transitions from the L classes to C$-$ are possible but much rarer~\cite{2015PhRvD..92f4016G}. 

The $\Delta\Phi$  morphologies are also intimately related to the spin--orbit resonances described in Sec.~\ref{secbounds}. These are the locations $\kappa_\pm$ where $\delta\chi_-=\delta\chi_+$. One can show \cite{2015PhRvD..92f4016G} that $\Delta\Phi=\pi$ at $\kappa_-$ and $\Delta\Phi=0$ at $\kappa_+$.  The resonances are therefore the limits of the two librating morphologies when the libration amplitude goes to zero. A more careful investigation of this limit with a formal Taylor expansion is an interesting avenue for future work.

\subsection{Up--down instability}

The spin-aligned configurations first mentioned in Sec.~\ref{lookfor}, where $\cos\theta_{1,2}=\pm 1$, are all equilibrium configurations of the equations of motion (\ref{dS1st})-(\ref{OmegaL}) ---though not all of them are stable.
Reference~\cite{2015PhRvL.115n1102G} first showed that up --down binaries ---in which the primary BH spin is aligned with the orbital angular momentum ($\theta_1=0$) and the secondary BH spin is antialigned ($\theta_2=\pi$)--- are unstable to spin precession beyond a critical orbital separation in their inspirals~\cite{2016PhRvD..93l4074L,2020PhRvD.101l4037M,2021PhRvD.103f4003V}. In such cases, perturbations to the spin directions leads to wide precession cycles rather than small-amplitude oscillations about alignment. Reference~\cite{2020PhRvD.101l4037M} further showed that unstable up--down sources asymptote to a well-defined and predictable endpoint at small separations, rather than dispersing in the available parameter space.

Here we reinvestigate these results using the new parametrization of the dynamics in terms of $\delta\chi$.
Let $\delta\chi^*$ denote the spin parameter for an aligned-spin configuration and $\alpha_i = \cos\theta_i = \pm1$ denote the spin--orbit alignment of the two BHs ($i=1,2$). For the four aligned-spin configurations one has
\begin{align}
\delta\chi^*
&=
\frac{\chi_1\alpha_1 - q\chi_2\alpha_2}{1+q}
\, , \\
\chi_\mathrm{eff}^*
&=
\frac{\chi_1\alpha_1 + q\chi_2\alpha_2}{1+q}
\, , \\
\frac{\kappa^*}{M^2}
&=\frac{1}{2q(1+q)^2}
\Bigg[2q(\chi_1\alpha_1+q^2\chi_2\alpha_2) +
\notag \\ &+
(\chi_1^2 + q^4\chi_2^2 + 2q^2(1+q)^2\chi_1\chi_2\alpha_1\alpha_2) \sqrt{\frac{M}{r}}
 \Bigg] 
\, .
\end{align}
Taking a second time derivative in Eq.~(\ref{thirddegree}) and using Eqs.~(\ref{sigma3})-(\ref{sigma1}) we therefore have, to leading order in a perturbation $\delta\chi-\delta\chi^*$, that~\cite{2020PhRvD.101l4037M}
\begin{align}
\frac{\dd^2}{\dd t^2} (\delta\chi-\delta\chi^*) + \omega^2 (\delta\chi-\delta\chi^*)
\simeq
0
\, ,
\label{oscillator}
\end{align}
where %
\begin{align}
M^2 \omega^2(r) 
&=
-\bar{\sigma}(3\sigma_3\delta\chi^*+\sigma_2)
\\&=
\frac{9}{4} \left[ \left(\frac{1-q}{1+q}\right)^2\frac{r}{M} - 2\frac{1-q}{1+q}\delta\chi^*\sqrt{\frac{r}{M}} + {\chi_\mathrm{eff}^*}^2 \right]
\nonumber \\
&\times \left(\sqrt{\frac{r}{M}}-\chi_\mathrm{eff}^*\right)^2 \left(\frac{M}{r}\right)^7
\label{instabilityomega}
\end{align}
determines the oscillation frequency of the perturbed state. An instability occurs when this frequency becomes complex. Using Eq.~(\ref{cubicandroots}) and Vieta's formulas, $\omega=0$ corresponds to $3\delta\chi^* = \delta\chi_- + \delta\chi_+ + \delta\chi_3/(1-q)$ .%

 The repeated root $r=\chi_{\rm eff}^2 M$ from  Eq.~(\ref{instabilityomega}) is unphysical since $|\chi_\mathrm{eff}|\leq1$. The other two roots are
\begin{align}
\sqrt{\frac{r_\pm}{M}}
&=
\frac{1+q}{1-q} \left( \delta\chi^* \pm \sqrt{{\delta\chi^*}^2-{\chi_\mathrm{eff}^*}^2} \right)
 \\
&=
\frac{\chi_1\alpha_1 - q\chi_2\alpha_2 \pm 2\sqrt{-q\chi_1\chi_2\alpha_1\alpha_2}}{1-q}
\, .
\label{rplusminus}
\end{align}
It is
straightforward
to show that $r_\pm$ can only take real and physical values for $\alpha_1=-\alpha_2=1$, i.e., the up--down configuration.
In particular, unstable motion occurs at orbital separations $r_{\rm UD +} \geq r \geq r_{\rm UD -}$, where~\cite{2015PhRvL.115n1102G} 
\begin{align}
\frac{r_{\rm UD \pm}}{M} = \frac{(\sqrt{\chi_1}\pm\sqrt{q\chi_2})^4}{(1-q)^2}
\, .
\end{align}
From Eq.~(\ref{instabilityomega}) one has $\lim_{r \to \infty}\omega^2(r)\geq0$ such that $r_{\rm UD +}$ marks the onsets of unstable spin precession. We also note stability in the limits of equal masses ($q\to 1$), extreme mass ratios ($q\to 0$), and zero spins ($\chi_{1,2}\to 0$).

After the instability is triggered, up--down binaries do not disperse in parameter space but approach a well-defined endpoint. For $r>r_{\rm UD +}$, one can show that up--down binaries are in the $\kappa_+$ spin--orbit resonance \cite{2015PhRvL.115n1102G,2020PhRvD.101l4037M}. Since resonant binaries remain so~\cite{2020PhRvD.101l4037M}, the endpoint of the up--down instability can be found using Eq.~(\ref{deltakappa}) as the formal $r\to 0$ limit of binary configurations with $\kappa=\kappa_+$.

In this limit, the discriminant equation $\Delta=0$ can be solved analytically in $\kappa/u$ to obtain 
\begin{align}
&\lim_{r\to 0} \frac{\kappa_-}{u M^4} = \max\bigg[\frac{(\chi_1-q^2\chi_2)^2}{(1+q)^4},
\notag \\ &\qquad\qquad\qquad\quad\frac{1-q}{(1+q)^4} (\chi_1^2-q^3\chi_2^2) +  \frac{q\chi_\mathrm{eff}^2}{(1+q)^2} 
\bigg] \,,
\\
&\lim_{r\to 0} \frac{\kappa_+}{u M^4} = \frac{(\chi_1+q^2\chi_2)^2}{(1+q)^4}\,.
\end{align}
According to Eq.~(\ref{Svsdeltachi}), one has $\kappa\sqrt{r}\propto\kappa/u\to S^2$ as $r\to0$.  Using Eqs.~(\ref{costheta12angles}) and (\ref{chieff}), and considering that $\kappa_+$ implies $\Delta\Phi=0$ while $\kappa_-$ implies $\Delta\Phi=\pi$, we find

\begin{align}
\label{res0m1}
\lim_{r\to0, \kappa\to \kappa-} \!\!\!\!
 \cos\theta_1
&=
\begin{cases}
\displaystyle
\frac{1\!+\!q}{\chi_1\!-\!q\chi_2}\chi_\mathrm{eff}
\;\;\; \mathrm{if} \;\;\; |\chi_\mathrm{eff}| \leq \frac{|\chi_1\!-\!q\chi_2|}{1\!+\!q}
\, , \\[10pt]
\displaystyle
\frac{\chi_1^2-q^2 \chi_2^2+(1\!+\!q)^2
   \chi_{\rm eff}^2}{2 (1\!+\!q) \chi_1
   \chi_{\rm eff}}
\;\;\; \mathrm{otherwise}
\, ,
\end{cases}
\\
\label{res0m2}
\lim_{r\to0, \kappa\to \kappa-} \!\!\!\!
\cos\theta_2
&=
\begin{cases}
\displaystyle
\frac{1\!+\!q}{q\chi_2\!-\! \chi_1}\chi_\mathrm{eff}
\;\;\; \mathrm{if} \;\;\; |\chi_\mathrm{eff}| \leq \frac{|\chi_1\!-\!q\chi_2|}{1\!+\!q}
\, , \\[10pt]
\displaystyle
\frac{q^2 \chi_2^2 -\chi_1^2+(1\!+\!q)^2
   \chi_{\rm eff}^2}{2 q (1\!+\!q) \chi_2
   \chi_{\rm eff}}
\;\;\; \mathrm{otherwise}
\, ,
\end{cases}
\\
\label{res0p1}
\lim_{r\to0, \kappa\to \kappa+}\!\!\!\!
 \cos\theta_1  &=
\frac{1+q}{\chi_1+q\chi_2}\chi_\mathrm{eff}\,,
\\
\label{res0p2}
\lim_{r\to0, \kappa\to \kappa+} \!\!\!\! 
\cos\theta_2
&=
\frac{1+q}{\chi_1+q\chi_2}\chi_\mathrm{eff}
\, .
\end{align}
These are the generic limits of the two spin--orbit resonances as $r\to 0$. We arrive to the specific case of up--down binaries by setting $\chi_\mathrm{eff}=(\chi_1-q\chi_2)/(1+q)$ in Eqs.~(\ref{res0p1}-\ref{res0p2}). The endpoint of the up-down instability is a precessing configuration with \cite{2020PhRvD.101l4037M}
\begin{align}
\label{endpoint1}
\begin{dcases}
\cos\theta_1 =  \frac{\chi_1-q\chi_2}{\chi_1+q\chi_2}\,,
\\
\cos\theta_2 =  \frac{\chi_1-q\chi_2}{\chi_1+q\chi_2}\,,
\\
\Delta\Phi = 0\,.
\end{dcases}
\end{align}

Because of such unstable behavior, we find that binaries that cross the up--down instability onset are the most challenging to evolve numerically. This includes binaries with $\cos\theta_1=1$ and $\cos\theta_2=-1$ evolved forward in time as well as binaries close to the endpoint of Eq.~(\ref{endpoint1}) evolved backward in time. Numerical challenges related to up--down binaries were also reported for the independent implementation described in Ref.~\cite{2022PhRvD.106b3001J}.

\subsection{Wide nutation}

Reference~\cite{2019CQGra..36j5003G} showed that, under specific conditions, BH  spins can oscillate from full alignment to full anti-alignment within a single period $\tau$. This phenomenon, which we dubbed ``wide nutation,'' corresponds by definition to the largest possible nutational motion in BH binary dynamics. Hints of these configurations were previously found in Refs.~\cite{2015PhRvL.114n1101L,2016PhRvD..93d4031L}. 

During the inspiral of a BH binary, wide nutation can only occur for either the primary or the secondary BH, not both, and only if the orbital separation is smaller than the threshold 
\begin{align}
\frac{r_{\rm wide}}{M} =
\left( \frac{\chi_1-q\chi_2}{1-q} \right)^2  \,.
\end{align}
More specifically, the wide-nutation condition for the primary BH corresponds to the constraints $\cos\theta_1(\delta\chi_-)=-1$ and  $\cos\theta_1(\delta\chi_+)=+1$. These are
satisfied
if \cite{2019CQGra..36j5003G}
\begin{align}
 r&\leq r_{\rm wide}\,,
 \\
 \chi_1&\leq \chi_2\,,
 \\
 \chi_{\rm eff} &= 
 -\frac{1- q}{1+q} \sqrt{\frac{r}{M}}\,,
 \\
\frac{\kappa}{M^2} &= 
 \frac{\chi_1^2 - 2 q \chi_1^2 + q^4 \chi_2^2 - 
 2 q^2(1 - q)  (r/M) }{2 q (1 + q)^2 \sqrt{r/M}} \,.
\end{align}
For the secondary BH, the relevant conditions are $\cos\theta_2(\delta\chi_-)=1$ and  $\cos\theta_2(\delta\chi_+)=-1$ which can be translated to~\cite{2019CQGra..36j5003G}
\begin{align}
 r&\leq r_{\rm wide}\,,
 \\
 \chi_2&\leq \chi_1\,,
 \\
 \chi_{\rm eff} &= 
 \frac{1- q}{1+q} \sqrt{\frac{r}{M}}\,,
 \\
\frac{\kappa}{M^2}  &= 
 \frac{\chi_1^2 - 2 q^3 \chi_2^2 + q^4 \chi_2^2 + 
 2 q(1 - q)  (r/M) }{2 q (1 + q)^2 \sqrt{r/M}}  \,.
\end{align}

\subsection{Estimators: $\chi_{\rm p}$}

The most commonly used estimator to quantify spin precession in GW data is the so-called $\chi_{\rm p}$ parameter first introduced by \citeauthor{2015PhRvD..91b4043S}~\cite{2015PhRvD..91b4043S} 
\begin{align}
\chi_{\rm p}^{(\rm heu)} &= \max \left( \chi_1 \sin\theta_1, q \frac{3+4q}{4+3q} \chi_2 \sin\theta_2 \right)\,.
\label{chipgen}
\end{align}
This expression was shown to be inconsistent in Ref.~\cite{2021PhRvD.103f4067G} and generalized using a full timescale separation. Their amended definition reads 
\begin{align}
\chi_{\rm p} &=  \Bigg[ (\chi_1 \sin\theta_1)^2 + \left( q \frac{3+4q}{4+3q} \chi_2 \sin\theta_2 \right)^2
 \notag \\
  &\qquad +2 q \frac{3+4q}{4+3q} \,\chi_1 \chi_2  \sin\theta_1 \sin\theta_2 \cos\Delta\Phi \Bigg]^{1/2}\,
\label{chipgen}
\end{align}
and, crucially, includes all variations that take place on the precession timescale. It can thus be precession averaged without ambiguities at the PN order considered here (Sec.~\ref{precavrules}), resulting in a precession estimator $\langle \chi_{\rm p}\rangle$ that only varies on the radiation-reaction timescale; see Refs.~\cite{2021PhRvD.104h4002B,2022PhRvD.106h4040D,2022PhRvD.106j3013M,2022CQGra..39l5003H} for applications and Refs.~\cite{2020PhRvD.102d1302F,2021PhRvD.103h3022T,2021PhRvD.103l4026G} for alternative estimators. 

While it is straightforward to estimate $\langle \chi_{\rm p}\rangle$ numerically using our new formulation based on $\delta\chi$, we were not able to solve the resulting integral analytically. We note however that the root mean square $\sqrt{ \langle \chi_{\rm p}^2 \rangle}$ can instead be written down in closed form~\cite{2021arXiv210610291K}.
Using Eqs.~(\ref{ct1})-(\ref{cosdeltaphi}) we first write
\begin{align}
\label{chip2poly}
\chi_{\rm p}^2 &= \bar\lambda (\lambda_2 \delta\chi^2 +  \lambda_1 \delta\chi + \lambda_0)
\end{align}
where $\bar \lambda$ and $\lambda_{i}$ are coefficients that depend on $\kappa$, $\chi_{\rm eff}$, $r$, $q$, $\chi_1$, and $\chi_2$, as provided in Appendix~\ref{polycoeffs}. Using Eq.~(\ref{deltachitilde}) we obtain
\begin{align}
\sqrt{ \langle \chi_{\rm p}^2 \rangle }&= \sqrt{\bar\lambda} \big[
(\delta\chi_+ - \delta\chi_-)^2 \lambda_2  {\langle \delta\tilde\chi ^2\rangle} 
+ (\delta\chi_+ - \delta\chi_-)
\notag \\
&\!\!\!\!\!\times (\lambda_1 + 2 \delta\chi_- \lambda_2){\langle \delta\tilde\chi\rangle} 
+(\delta\chi_- \lambda_1 + \delta\chi_-^2 \lambda_2 + \lambda_0)  \big]^{1/2}\!,
\label{chiprmsss}
\end{align}
where  ${\langle \delta\tilde\chi \rangle}$ and ${\langle \delta\tilde\chi^2 \rangle}$ are given in Eqs.~(\ref{deltachitildeav}) and (\ref{deltachitildeav2}), respectively, using elliptic integrals %
(see  Fig.~\ref{mfunction}).

At infinitely large separations one has $\theta_{1,2}=$ const. and $\langle\cos\Delta\Phi \rangle= 0$ (Sec.~\ref{notable}). Plugging these into   Eq.~(\ref{chipgen})  and computing the average returns\footnote{Equation~(\ref{chipavlim})  is equivalent to Eq.~(19) in Ref.~\cite{2021PhRvD.103f4067G} but written in a more compact form.} 
\begin{align}
 \lim_{r\to \infty} \langle \chi_{\rm p} \rangle &= 
\frac{2}{\pi}\left( \chi_1 \sin\theta_1 + q \frac{3+4q}{4+3q} \chi_2 \sin\theta_2 \right) 
\notag\\
&\times E \Bigg[\left(4 q \frac{3+4q}{4+3q}\chi_1 \chi_2 \sin\theta_1\sin\theta_2\right) 
\label{chipavlim}
\notag \\ 
&\times \left( \chi_1 \sin\theta_1 + q \frac{3+4q}{4+3q} \chi_2 \sin\theta_2 \right)^{-2} \Bigg]
\end{align}
and
\begin{align}
 \lim_{r\to \infty}\sqrt{ \langle \chi_{\rm p}^2 \rangle } &= \sqrt{(\chi_1 \sin\theta_1)^2 + \left( q \frac{3+4q}{4+3q} \chi_2 \sin\theta_2 \right)^2}
\end{align}
(recall that $E$ is the complete elliptic integral of the second kind).
In the limit of single-spin binaries, both these expressions reduce to the ``heuristic'' definition $\chi_{\rm p}^{(\rm heu)}$ \cite{2015PhRvD..91b4043S} reported in Eq.~(\ref{chipgen}); see Ref.~\cite{2021PhRvD.103f4067G} for details.

It is important to stress that $\langle \chi_{\rm p}\rangle$ and  $\sqrt{ \langle \chi_{\rm p}^2 \rangle}$ are two different estimators and one is not an approximation of the other. 
In particular, one necessarily has $\sqrt{ \langle \chi_{\rm p}^2 \rangle} \geq \langle \chi_{\rm p}\rangle$ for any BH binary (this can be proven using the Cauchy-Schwarz inequality in the $L^2$ Hilbert space). The domain of both estimators is $[0,2]$, unlike the earlier definition of Ref.~\cite{2021PhRvD.103f4067G} which is defined in $[0,1]$. The additional region $[1,2]$ is unique to binaries with two precessing spins and can be exploited to probe the underlying physics~\cite{2022PhRvD.106h4040D}.

\begin{figure}
\includegraphics[width=\columnwidth]{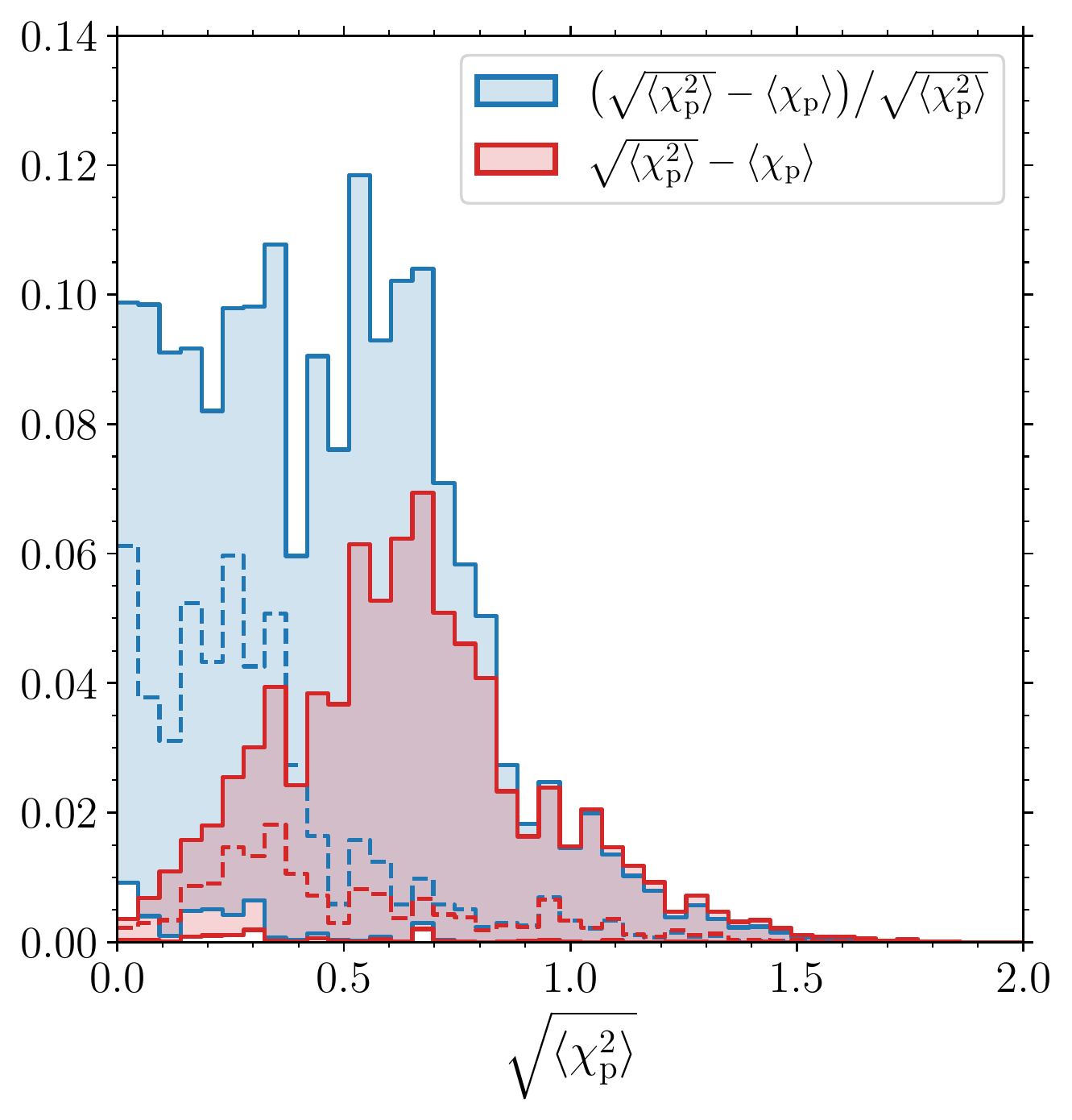}
\caption{Relative (blue) and absolute (red) differences between the two precession estimators $\langle \chi_{\rm p}\rangle$ and  $\sqrt{ \langle \chi_{\rm p}^2 \rangle}$. We consider a sample of BH binaries where $\sqrt{ \langle \chi_{\rm p}^2 \rangle}$ is uniformly distributed in $[0,2]$, as obtained from reweighting a base distribution where $q$, $\chi_1$, and $\chi_2$ are uniformly distributed  in $[0.1,1]$, $r=10M$, and  spin directions are isotropic. Shaded areas encompass $90\%$ of the binaries in each bin. Dashed lines mark the median error values. 
}
\label{chiprmsfig}
\end{figure}

Figure~\ref{chiprmsfig} shows the relative and absolute difference between  $\langle \chi_{\rm p}\rangle$ and  $\sqrt{ \langle \chi_{\rm p}^2 \rangle}$. For this exercise, we generated a sample of BH binaries with uniform values of $\sqrt{ \langle \chi_{\rm p}^2 \rangle}\in[0,2]$. This was obtained by numerically reweighting an initial distribution where $q$, $\chi_1$, and $\chi_2$ are distributed uniformly in $[0.1,1]$,  the separation is kept fixed to $r=10M$, and spin directions are isotropic. We find that the largest relative (absolute) differences between the two are $\ssim 0.12$ ($\ssim 0.07$) and take place for values of $\sqrt{ \langle \chi_{\rm p}^2 \rangle}\ssim 0.7$.

The crucial  difference between the two is that $\sqrt{ \langle \chi_{\rm p}^2 \rangle}$ is about 
$10^4$ times faster to evaluate than $\langle \chi_{\rm p}\rangle$. For the sources in Fig.~\ref{chiprmsfig}, the average computational time on a standard off-the-shelf laptop was $\ssim 0.1$ s for $\langle \chi_{\rm p}\rangle$ and $\ssim 10^{-5}$ s for  $\sqrt{ \langle \chi_{\rm p}^2 \rangle}$. This considerable speedup might turn out to be useful when exploring the two-spin generalization of $\chi_{\rm p}$ for sampling purposes in GW parameter-estimation pipelines (cf. Ref.~\cite{2022arXiv221007912W} for current attempts using \textsc{rift}).

\subsection{Estimators: Five phenomenological parameters}

A more phenomenological approach to quantify the amount of spin precession in a BH binary includes exploiting frequencies and angles that are directly related to the timescale separation of Eq.~(\ref{hierarchy})~\cite{2021PhRvD.103l4026G,2021MNRAS.501.2531S,2022PhRvD.106b4019G}. 
Considering the joint precessional and nutational motion of $\boldsymbol{L}$ about $\boldsymbol{J}$, we define the following ``geometric'' estimators \cite{2021PhRvD.103l4026G}:
\begin{enumerate}
    \item the precession amplitude $\langle\theta_L\rangle$, which is the average of Eq.~(\ref{thetaL});
    \item the precession frequency $\langle \Omega_L \rangle =\alpha/\tau$, which is the average of Eq.~(\ref{thisomegaL});
    \item the nutation amplitude $\Delta \theta_L =  \left[\theta_L (\delta \chi_{-}) - \theta_L (\delta \chi_{+} )\right]/2$ describing the variation of $\theta_L$ during a cycle;\footnote{The signs of $\Delta \theta_L$ and $\Delta \Omega_L$ as reported here are chosen for consistency with Ref.~\cite{2021PhRvD.103l4026G}.} %
    \item the nutation frequency $\omega = {2\pi}/{\tau}$ from Eq.~(\ref{tauperiod});
    \item the variation of the precession frequency due to nutation, $\Delta \Omega_L= \left[\Omega_L (\delta \chi_{-}) - \Omega_L (\delta \chi_{+} )\right]/2$. %
\end{enumerate}

In the case where a single spin dominates the binary dynamics, the polar motion is suppressed and the nutational parameters $\Delta \theta_L$ and $\Delta \Omega_L$ become irrelevant. Similarly, binaries do not nutate when $q = 1$ and in the case of the spin--orbit resonances (because $S$ is constant, see Ref.~\cite{2021PhRvD.103l4026G} for details). %

\section{Interface with other investigations}
\label{interface}

\subsection{Orbit-averaged inspirals}

The precession-averaged approach is built on top of the standard orbit-averaged formulation of the BH binary dynamics, which is briefly presented here for completeness. In short, an orbit-averaged PN integration requires solving nine coupled ODEs given in Eqs.~(\ref{dS1st})-(\ref{OmegaL}) for $\boldsymbol{L}(t),\boldsymbol{S}_1(t)$ and $\boldsymbol{S}_2(t)$.
Equation~(\ref{dLdtorb}) needs to be supplemented with a prescription for $\dd L/\dd t$ which encodes radiation reaction. The momentum flux including (non-) spinning terms up to (3.5PN) 2PN reads (see Refs.~\cite{2008PhRvD..78d4021R, 2009PhRvD..80d4010R,2010PhRvD..81h4054K} and references therein)
\begin{widetext}
\begin{align}
&\frac{\dd L}{\dd t}= -\frac{32}{5} \frac{M^8}{L^7} \Bigg\{ 1-
\frac{M^4}{L^2} \eta^2 \frac{743 +924 \eta }{336}
+
\frac{M^6}{L^3} \eta^3  \Bigg[ 4 \pi 
- \eta \chi_1\cos\theta_1 \left(\frac{113}{12 q } + \frac{25}{4} \right)
- \eta \chi_2\cos\theta_2\left(\frac{113q}{12} + \frac{25}{4} \right)
\Bigg]
+  \frac{M^8}{L^4} \eta^4
\notag\\&
\times 
\Bigg[\frac{34103}{18144} + \frac{13661}{2016}\eta 
 +\frac{59}{18}\eta^2
 + \frac{\eta \chi_1^2}{96 q} ( 719 \cos^2\theta_1
 - 233 ) 
+ \frac{\eta q\chi_2^2}{96} \left( 719 \cos^2\theta_2 - 233 \right)  
+ \frac{\eta \chi_1 \chi_2}{48} 
( 474 \cos\theta_1 \cos\theta_2 
\notag\\&
- 247 \cos\Delta\Phi \sin\theta_1\sin\theta_2 )
\Bigg]
-
 \frac{M^{10}}{L^5} \eta^5 \pi \frac{4159 +15876 \eta}{672} 
+  \frac{M^{12}}{L^6} \eta^6\Bigg[\frac{16447322263}{139708800}
+\frac{16}{3}\pi^2 -\frac{1712}{105}\bigg(\gamma_E +\ln\frac{4\eta M^2}{L}\bigg)
\notag\\& 
+\left( \frac{451}{48} \pi^2 -  \frac{56198689}{217728} \right) \eta
+ \frac{541}{896}\eta^2 -\frac{5605}{2592}\eta^3\Bigg]
 +   \frac{M^{14}}{L^7} \eta^7 \pi \Bigg[ -\frac{4415}{4032} + \frac{358675}{6048}\eta +\frac{91495}{1512}\eta^2 \Bigg]\Bigg\}
\,,
\label{radiationreaction}
\end{align}
\end{widetext}
where $\eta=q/(1+q)^2$ and $\gamma_E\simeq 0.577$ is  Euler’s  constant. In particular, spins enter at 1.5PN, which is the reason why the precession-averaged approach to radiation reaction presented in Sec.~\ref{precavinspiral} can only be extended up to 1PN order. Orbit-averaged and precession-averaged integrations have been extensively compared against each other in Ref.~\cite{2015PhRvD..92f4016G}.

\subsection{Comparing to GW measurements}

GW measurements are usually provided in the form of samples from a posterior distribution, where the spin directions are quoted at a reference emission frequency $f_{\rm GW}$. For LIGO/Virgo, this is often (but not always~\cite{2020PhRvL.125j1102A}) set to $20$ Hz~\cite{2019PhRvX...9c1040A,2021PhRvX..11b1053A,2021arXiv211103606T}.
When interpreting GW data in light of our formalism, one needs to convert $f_{\rm GW}$ into a PN separation $r$. To this end, let us first write the dimensionless orbital frequency
\begin{align}
\tilde\omega=  \frac{\pi  G}{ c^3} M f_{\rm GW} \simeq 1.48 \times 10^{-5} \left(\frac{M}{M_\odot}\right) \left(\frac{f_{\rm GW}}{{\rm Hz}}\right)\,,
\end{align}
where we reinstated physical units for clarity.
 At 2PN, the conversion between $\tilde\omega$ and $r$ is given by~\cite{1995PhRvD..52..821K}
\begin{align}
\frac{r}{M}&= \frac{1}{\tilde\omega^{2/3}} \bigg\{ 1 - \tilde\omega^{2/3} \left[ 1-  \frac{q}{3 (1 + q)^2} \right] 
\notag \\
& - \frac{\tilde\omega}{3 (1+q)^2}
\left[(2 + 3 q) \chi_1 \cos\theta_1+ 
  q (3 + 2 q) \chi_2 \cos\theta_2\right]
\notag \\
& +\tilde\omega^{4/3}
\frac{q }{2 (1 + q)^2} \bigg[ \frac{19}{2} 
+ \frac{2 q}{9 (1 + q)^2}  
\notag \\ 
 &+  \chi_1\chi_2 \left( 2 \cos\theta_1 \cos\theta_2 - 
\cos\Delta\Phi \sin\theta_1\sin\theta_2 \right)\bigg] \bigg\}\,,
\end{align}
while the inverse transformation is~\cite{1995PhRvD..52..821K} %
\begin{align}
\tilde\omega &= \left(\frac{M}{r}\right)^{3/2} \bigg\{ 1- \frac{M}{r} \left[ 3- \frac{q}{(1+q)^2}\right]
\notag \\
& - \left(\frac{M}{r}\right)^{\!\!3/2}  
\!\!\!\!\!\frac{1}{(1\!+\!q)^2}
\left[(2 \!+\! 3 q) \chi_1 \cos\theta_1+ 
  q (3 \!+\! 2 q) \chi_2 \cos\theta_2\right]
\notag \\
& +\left(\frac{M}{r}\right)^{2}  \bigg[ 
6 + \frac{41 q}{4 (1 + q)^2} +  \frac{q^2}{(1 + q)^4}
+
\frac{3 q }{2 (1\! + \!q)^2}  
\notag \\
& \times \chi_1\chi_2 \left( 2 \cos\theta_1 \cos\theta_2 - 
\cos\Delta\Phi \sin\theta_1\sin\theta_2 \right)\bigg] \bigg\}^{1/2}.
\end{align}

For current LIGO/Virgo observations of BH binaries, $f_{\rm GW}$ is only a few orbits away from merger, where $t_{\rm pre}$ is likely to be comparable with $t_{\rm rad}$ and the precession-averaged formalism breaks down \cite{2015PhRvD..92f4016G,2022PhRvD.106b3001J,2022PhRvD.106b3001J}. A more solid approach, therefore, is that of a {hybrid} evolution~\cite{2016PhRvD..93l4066G,2020CQGra..37v5005R}, where one first propagates LIGO samples backward using an orbit-averaged integration (which is numerically expensive but keeps track of the precession phase) and then switches to a precession-averaged formulation (which is less accurate but can be easily be extended all the way to $r=\infty$, i.e., $f_{\rm GW}=0$).

The transition threshold $r_{\rm t}$ between the orbit-averaged and the precession-averaged approach is somewhat arbitrary, but it has a negligible effect as long as it falls in the regime where $t_{\rm pre}\ll t_{\rm rad}$. For a rule of thumb, we find that switching between the two formulations at $r_{\rm t}=1000 M$ provides accurate results while maintaining the computational cost under control~\cite{2015PhRvD..92f4016G,2020CQGra..37v5005R}. For a deeper investigation of the transition between orbit- and precession-averaged PN integrations see Ref.~\cite{2022PhRvD.106b3001J}.

\subsection{Remnant properties}

Modeling the properties of remnant BHs left behind following binary mergers has important applications in both astrophysics and GW analyses (cf. Refs.~\cite{2008ApJ...684..822B,2019MNRAS.482.2991A,2020CQGra..37v5005R,2021NatAs...5..749G,2021ApJ...915L..35K} for a few, non-exhaustive examples). 
In line with previous versions of \textsc{precession}~\cite{2016PhRvD..93l4066G}, the new code presented in Sec.~\ref{pythonsec} includes fitting formulas to numerical-relativity simulations that model the mass $M_\mathrm{f}$, spin $\chi_\mathrm{f}$, and proper velocity (or kick) $v_\mathrm{f}$ of the post-merger remnant. In particular, we implement phenomenological expressions from Ref.~\cite{2012ApJ...758...63B} for the final mass, Ref.~\cite{2016ApJ...825L..19H} for the final spin, and Ref.~\cite{2016PhRvD..93l4066G} for the BH kick. These were assembled using several NR simulations available at the time (see references therein for details and credits to the various NR runs). The direction of the final spin is approximated using the total angular momentum before merger \cite{2016ApJ...825L..19H}.

Those formulas provide estimates of $M_\mathrm{f}$, $\chi_\mathrm{f}$, $v_\mathrm{f}$ as a function of $q$, $\chi_1$, $\chi_2$, $\theta_1$, $\theta_2$, and $\Delta\Phi$. More specifically, the final-mass prescription we implement does not depend on $\Delta\Phi$ while the kick velocity has an additional dependence on the orbital phase, which we assume to be randomly distributed~\cite{2016PhRvD..93l4066G}. Crucially, these predictions are inherently ill posed because they do not depend on the orbital separation, even though $r$ is a necessary coordinate to specify the binary configuration (see Sec.~\ref{lookfor}). The rationale is that those expressions should only be applied sufficiently close to merger ($r\simeq 10M$) where $t_{\rm pre}\ssim t_{\rm rad}$ and the spins do not precess much. 

A more accurate approach to predicting the properties of post-merger BHs relies on
surrogate modeling techniques
first developed for waveform approximants~\cite{2019PhRvL.122a1101V,2019PhRvR...1c3015V}. Remnant surrogates are data-driven fits to NR simulations that do not assume a specific functional form. While this solves the quoted ambiguity on the orbital separation, their predictions are limited to the region of the parameter space where NR coverage is sufficiently dense. Within their regime of validity, surrogate remnants are more accurate than the simple expressions implemented in \textsc{precession} and should be used~\cite{2019PhRvL.122a1101V}.

\section{Numerical implementation}
\label{pythonsec}

\subsection{Distribution and documentation}

A public implementation of our findings is distributed in the {\sc precession} module for the Python programming language. Version v1 of {\sc precession} was illustrated in Ref.~\cite{2016PhRvD..93l4066G}. The code presented here is tagged v2 and has been rewritten from scratch. In particular, we broke backward compatibility because the mathematical formulation presented in this paper could not be encapsulated into the existing routines.

The source code is available at  \cite{repo}
\begin{quote}
$\!\!$\href{https://github.com/dgerosa/precession}{github.com/dgerosa/precession}~~(source code).
\end{quote}
The code documentation can be browsed at  
\begin{quote}
$\!\!$\href{https://dgerosa.github.io/precession}{dgerosa.github.io/precession}~~(documentation)
\end{quote}
and includes a detailed list of all functions together with tutorials to perform some of the key operations. The code is distributed via the Python Package Index (PyPI) and can be installed with
\begin{quote}
$\!\!$\texttt{pip install precession}~~(installation).
\end{quote}
Dependencies are limited to {\sc numpy} \cite{2020Natur.585..357H} and {\sc scipy} \cite{2020NatMe..17..261V}. 

The general structure of the code is that of a toolbox, namely a series of functions that can be chained by the user to perform the desired calculation. In particular, we provide tools to (i)~capture the BH dynamics on the spin-precession timescale in closed form, (ii)~average generic quantities over a precession period, (iii)~numerically integrate the BH binary inspiral using both orbit- and precession-averaged approximations, (v)~evaluate spin-precession estimators, and (vi)~estimate the remnant properties. 

Code units are such that $G=c=M=1$, where $M$ is the total mass of the binary. There are a few exceptions where we interface our scale-free calculations with GW detectors. In those cases, inputs and outputs are more conveniently expressed in $M_\odot$, Hz, etc. The order of inputs and outputs respects the timescale hierarchy of precessing BH binaries, with variables varying on $t_{\rm pre}$ listed first, then those varying on $t_{\rm rad}$, and finally the constants of motion.

Vectorization via {\sc numpy} arrays is implemented whenever it is compatible with the adopted  {\sc numpy} and {\sc scipy} routines. For the case of polynomial root finding, we developed our own generalization as presented in Appendix ~\ref{rootsalg}. {\sc precession} functions can digest  inputs under the form of {\sc numpy}  arrays and perform operations on an element-by-element basis, in line with the {\sc numpy} broadcasting rules. By convention, outputs are returned as arrays of shape $(M,N)$, where $M$ is the number of features and $N$ is the number of binaries under study (as given in the input arrays). For consistency, this convention also applies to $N=1$ such that studying a single GW source returns two-dimensional arrays of shape $(M,1)$ and not one-dimensional arrays of length $M$ (this is somewhat inspired by the convention adopted in the popular {\sc scikit-learn} Python package~\cite{2011JMLR...12.2825P}).

Lengthy equations have been generated using the computer-algebra software Mathematica and exported to Python. Our source Mathematica notebook is made available in the {\sc precession} repository \cite{repo}.

\subsection{Performance}

\begin{figure}
\includegraphics[width=\columnwidth]{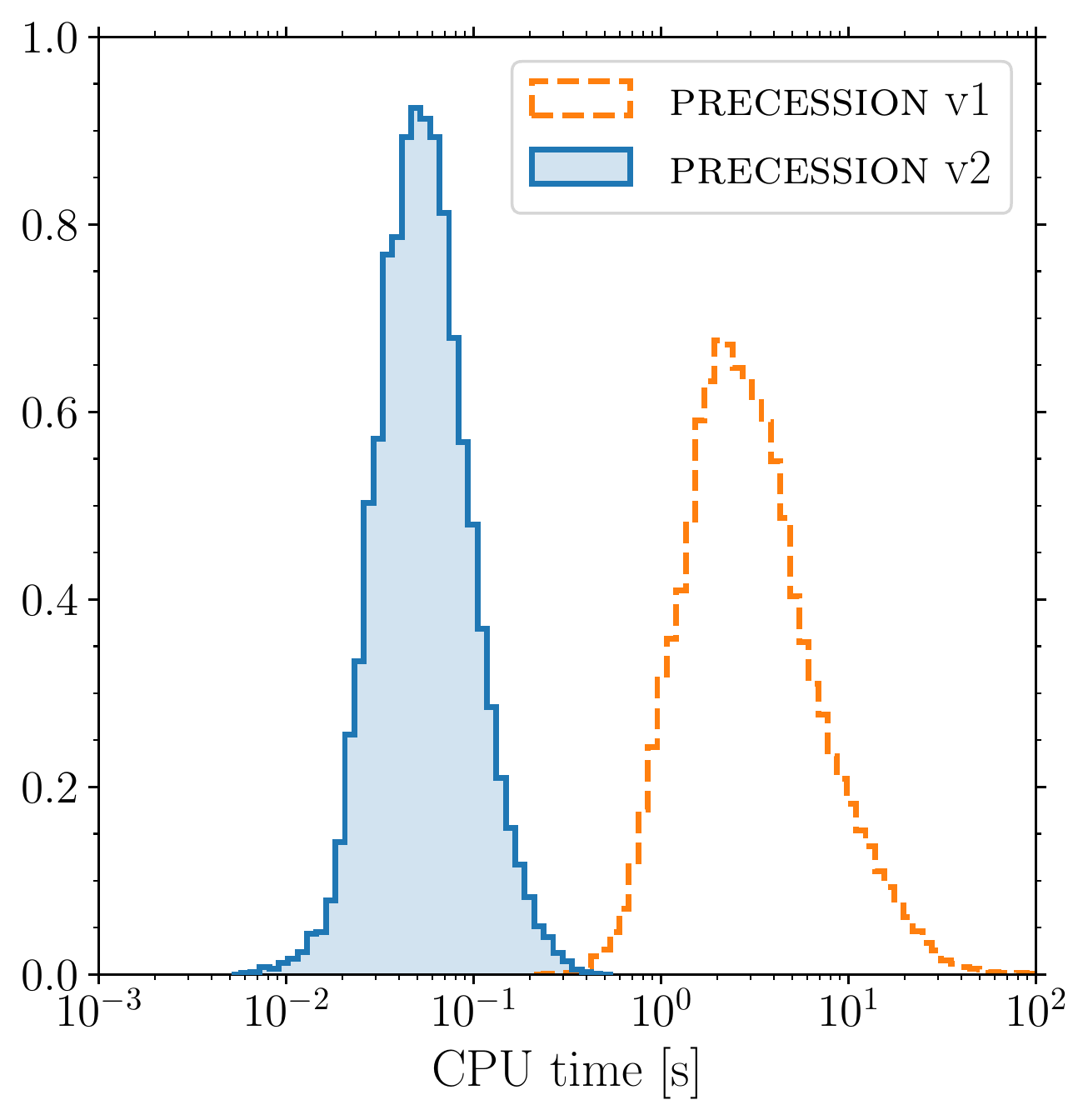}
\caption{Distribution of the CPU time (in seconds) required to perform precession-averaged evolutions. The blue histogram %
reports timings obtained with the version of the code presented in this paper (\textsc{precession} v2). The orange histogram %
reports timings obtained for the same BH binaries evolved with the version of the code presented in  Ref.~\cite{2016PhRvD..93l4066G} (\textsc{precession} v1). The unit on the y-axis is arbitrary. %
}
\label{p1vsp2}
\end{figure}

\begin{figure}
\includegraphics[width=\columnwidth]{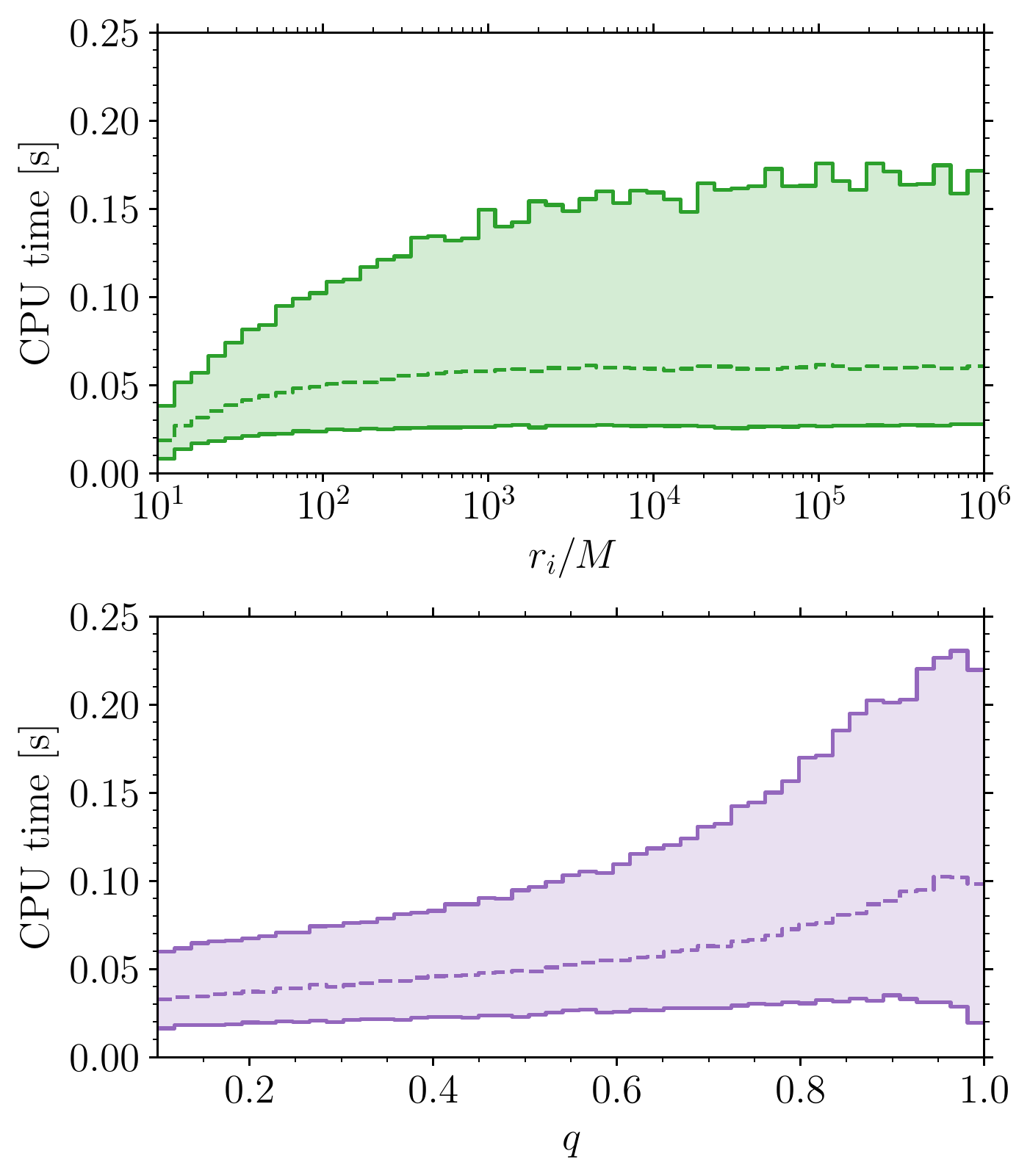}
\caption{CPU time (in seconds) required to perform precession-averaged evolutions in bins of initial separation $r_i$ (top panel, green) and mass ratio $q$ (bottom panel, purple). Dashed (solid) lines indicate the median  (90\% interval) wall-clock time recorded across a broad population of sources. %
}
\label{timepars}
\end{figure}

We test the performance of our new implementation on a population of $10^5$ BH binaries with $q$, $\chi_1$, and $\chi_2$ distributed uniformly in $[0.1,1]$ and isotropic spin orientations. We evolve these sources along their precession-averaged inspiral from $r_i$ to $r_f=10M$, where $r_i\in[10^6 M, 10M)$ is distributed uniformly in $\log r_i$. We record the execution times required to perform the entire procedure outlined in Sec.~\ref{bininsp}, i.e., both an integration of the $\dd\kappa/\dd u$ ODE as well as a resampling of the precessional phase at $r_f$. Tests were run on parallel threads using two Intel Xeon Gold 5220R processors.

Figure~\ref{p1vsp2} compares the performance of \textsc{precession} v2 against that of \textsc{precession} v1 from  Ref.~\cite{2016PhRvD..93l4066G}. We report wall-clock times 
$t_{\rm v2} = 0.06_{-0.03}^{+0.09}\,{\rm s}$ for the new code compared to  $t_{\rm v1} = 2.75_{-1.86}^{+10.38}\, {\rm s}$ obtained with the previous version  (where we quote medians and the 90\% interval across all simulated sources). This corresponds to a speedup of $t_{\rm v1}/t_{\rm v2}=  49.6_{-19.5}^{+104.4}$.

Figure~\ref{timepars} shows the execution times of the new code in bins of $r_i$ and $q$. The scaling with $r_i$ is essentially constant (or, more conservatively, logarithmic \cite{2015PhRvD..92f4016G}). Binaries with mass ratios close to unity take, on average, about a factor of $\lesssim 3$ longer to evolve compared to sources with $q\ssim 0.1$. This is expected because the importance of spin--spin couplings scales as $S_2/S_1\propto q^2$.

\subsection{Profiling}
\label{secprof}

\begin{figure*}
\includegraphics[width=\textwidth]{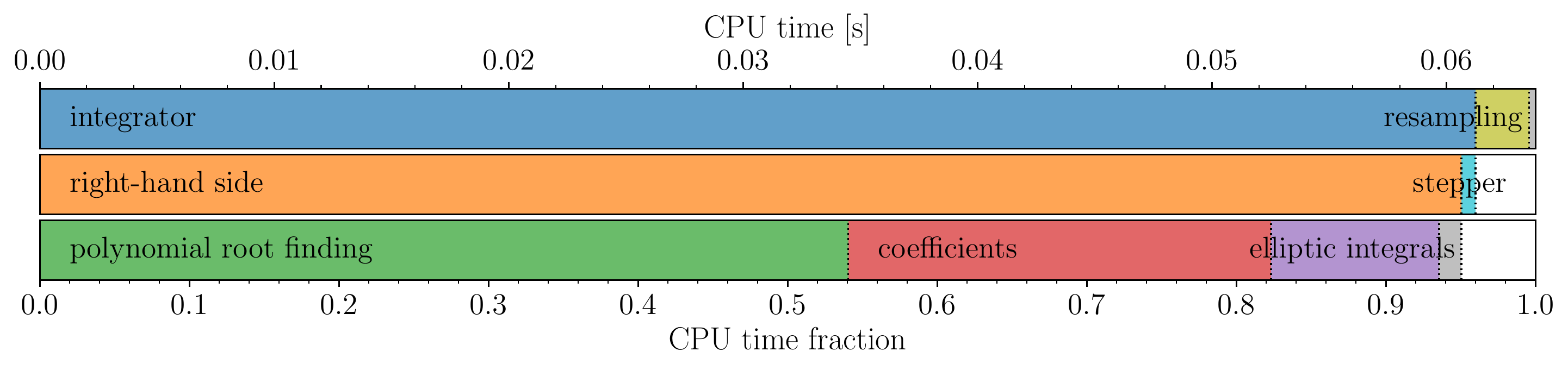}
\caption{Code profiling of \textsc{precession} v2. The length of each colored bar indicates the fraction of the total CPU time spent on a given operation, as indicated on the bottom x-axis. The top x-axis is rescaled to the mean  CPU time per binary $\ssim 0.064$ s. Performing a precession-averaged evolution requires an ODE integration (blue) and a resampling of the precessional phase at the final separation (yellow). In turn, the integrator requires multiple evaluations of the right-hand side (orange) and the ODE stepper (cyan). In turn, evaluating the right-hand side requires finding the roots of the cubic polynomial $\Sigma(\delta\chi)$ (green), computing the coefficients $\sigma_i$ (red), and evaluating elliptic integrals (purple). Minor additional operations are marked in gray. %
}
\label{homemadeprofiling}

\end{figure*}

Figure~\ref{homemadeprofiling} shows code profiling results for a set of precession-averaged evolutions from the same population described above. In particular, the ODE integrator takes about 96\% of the time while the remaining $\ssim 4\%$ is spent resampling the precessional phase at the final orbital separation. Within the integrator, the evaluation of the right-hand side from Eq.~(\ref{Svsdeltachiav}) requires the vast majority of the resources (about 95\% of the total execution time) while a minor fraction of the time is taken by the ODE stepper.  Deeper into the code, the computation of the right-hand side requires three operations with a noticeable computational footprint: the evaluation of the coefficients from Eqs.~(\ref{sigma3})-(\ref{sigma0}), the root finder to evaluate $\delta\chi_{\pm,3}$, and the evaluation of the elliptic integral from Eq.~(\ref{deltachitildeav}). These tasks require $\ssim 28\%$, $\ssim 54\%$, and $\ssim 11\%$ of the total execution time. Possible computational improvements include exploring just-in-time compilation for the evaluation of $\sigma_i$ \cite{2015llvm.confE...1L} as well as porting the polynomial root finder to GPUs~\cite{GHIDOUCHE201746}.

\section{Summary and future developments}
\label{concl}

The dynamics of precessing BH binaries is rich and fascinating. This paper presents a complete reinvestigation of the related phenomenology using multi-timescale methods. Our strategy relies on double averaging the equations of motion over both the orbital and the precessional timescale. Radiation reaction is then captured in a quasi-adiabatic fashion. 

Our previous approach \cite{2015PhRvD..92f4016G} parametrized the dynamics on the precession timescale using the magnitude of the total spin $S$. While intuitive, it results in a coordinate singularity when the two BHs have equal masses. On the other hand, the formulation presented here uses the weighted spin difference $\delta\chi$ which allows us to capture the $q \to 1$ limit, at least for finite orbital separations. The joint limits of $q\to 1$ and $r\to \infty$ still need to be fully understood, and we anticipate the solution will require identifying a new radiation-timescale parameter to be adopted instead of $\kappa$.

Using the new $\delta\chi$ formulation, we expanded upon previous results (most notably Ref.~\cite{2021arXiv210610291K}) and expressed the 2PN spin-precession dynamics in closed form. While some of the mathematical expressions presented in this paper might appear convoluted, the entire evolution on $t_{\rm pre}$ is written down in terms of elliptic integrals and Jacobi elliptic functions, which are extremely fast to evaluate using standard numerical libraries.

Our numerical implementation is distributed in v2 of the \textsc{precession} module for the Python programming language; see \href{https://github.com/dgerosa/precession}{github.com/dgerosa/precession} \cite{repo}. Performing precession-averaged binary BH inspirals from (infinitely) large separations to the PN breakdown takes $\lesssim 0.1 $s on a standard, off-the-shelf chip. %
This increased speedup has important applications in GW astronomy, including:
\begin{itemize}
\item[(i)] Post-processing long posterior chains describing GW events. These are provided at separations where BHs are visible and need to be propagated backward to separations where they form (e.g., Refs.~\cite{2022PhRvD.105b4076M,2022PhRvD.106b3001J}).
\item[(ii)] Evolve outputs from population-synthesis predictions of astrophysical nature. These are provided where BHs form and might need to be propagated forward to small separations where they become detectable (e.g., Refs.~\cite{2016ApJ...832L...2R,2018PhRvD..98h4036G}).
\end{itemize}

In this paper, we only tackled BHs on quasi-circular orbits. A generalization of our formalism to eccentric systems is under development; cf.  Refs.~\cite{2019PhRvD.100l4008P,2020PhRvD.102l3009Y,2021arXiv210610291K} for existing investigations. Further extensions include considering higher-order PN terms, as well as neutron stars
(or exotic compact objects)
in addition to BHs~\cite{2022arXiv221204657L,
2021PhRvD.103h4024K,2021PhRvD.103h4025K}.
These two lines of research might require a similar mathematical formalism as they both cause variations of $\chi_{\rm eff}$, which ceases to be a constant of motion. The dynamics presented in this paper
could provide the background solution for a perturbative approach where $\chi_{\rm eff}$ is allowed to undergo small oscillations. Finally, some of the most recent advances in PN theory include the identification of constants of motion for the non-averaged problem, without recursing to any adiabatic approximation~\cite{2017PhRvD..95j4054G,2021PhRvD.103f4066T,2023PhRvD.107j3040T,2022arXiv221001605S}. A detailed comparison of our predictions against theirs is another promising avenue for future work.

\acknowledgements
We thank Antoine Klein, Leo Stein, Ulrich Sperhake, Emanuele Berti, Mike Kesden,  Richard O'Shaughnessy, and Chad Henshaw for discussions.
We thank all users of {\sc precession} for their feedback. We thank the developers of {\sc scipy} pull requests No. 10504, No. 15135, and No.10304 for their implementation of Carlson elliptic integrals. D.Ger., G.F., M.M, D.Gan., and V.D.R. are supported by ERC Starting Grant No.~945155--GWmining, Cariplo Foundation Grant No.~2021-0555, MUR PRIN Grant No.~2022-Z9X4XS, and the ICSC National Research Centre funded by NextGenerationEU. D.Ger. is supported by Leverhulme Trust Grant No. RPG-2019-350. D.P.M. is supported by the International REU program at the University of Florida. D. Gan. is supported by the ``Study in Italy'' program of the Italian Ministry of Foreign Affairs and International Cooperation. Computational work was performed at CINECA with allocations through INFN, Bicocca, and ISCRA Project No. HP10BEQ9JB.

\appendix

\section{Polynomial coefficients}
\label{polycoeffs}

In this Appendix we list some lengthy equations that were omitted from the main body of the paper. %

We first report the coefficients entering the polynomial 
\begin{align}
\Sigma =\bar{\sigma} \sum_{i=0}^3 \sigma_i\delta\chi^i
\end{align}
of Eq.~(\ref{thirddegree}). These are:
\begin{align}
\bar\sigma &= 4608 \,  u^{10} M^{20}  \frac{  q^{11} }{(1 + q)^{27}}  \left[(1 + q)^2 - 2 u M^2 q \chi_{\rm eff}  \right]^2\,,
\label{sigmabar} \\
\sigma_3 &=  u M^2 (1-q)\,,
\label{sigma3} \\
\sigma_2 &=
-\frac{(1-q)^2 (1+q)}{2 q}-u M^2(1+q) \left(2 \frac{\kappa}{M^2} -\chi_{\rm eff}\right)
\notag \\
&
+2 u^2 M^4 \frac{  (1-q)}{(1+q)^3} (\chi_1^2-q^3 \chi_2^2)
\label{sigma2} \\
\sigma_1&=
\frac{(1-q) (1+q)^2 }{q} \left(2 \frac{\kappa}{M^2} -\chi_{\rm eff}\right)
\notag \\
& - uM^2 \frac{ 1-q}{q (1+q)^2} \left[2 (\chi_1^2+q^4 \chi_2^2)+q (1+q)^2 \chi_{\rm eff}^2\right]
\notag \\
&
+4 u^2 M^4\frac{q }{(1+q)^3} (\chi_1^2-q^2 \chi_2^2) \chi_{\rm eff}\,,
\label{sigma1} 
\end{align}
\begin{align}
\sigma_0 &= -\frac{(1+q)^3}{2 q} \left(2 \frac{\kappa}{M^2} -\chi_{\rm eff}\right)^2
\notag \\
&
+\frac{u  M^2}{q (1\!+\!q)} \left(2 \frac{\kappa}{M^2} -\chi_{\rm eff}\right) \left[2 (\chi_1^2\!+\!q^4 \chi_2^2)+q (1\!+\!q)^2\chi_{\rm eff}^2\right] 
\notag \\
& - \frac{2 u^2 M^4}{q (1+q)^5}  \left[(\chi_1^2-q^4\chi_2^2)^2\!+q (1+q)^3 (\chi_1^2+q^3 \chi_2^2) \chi_{\rm eff}^2\right]\!.
\label{sigma0}
\end{align}

We now list the coefficients entering the discriminant in Eq.~(\ref{deltakappa}). We break the calculation down as follows:
\begin{align}
\Delta = \bar\delta \sum_{i=0}^5 \delta_i \left(\frac{\kappa}{M^2}\right)^i = \bar\delta \sum_{i=0}^5 \sum_{j=0}^6 \delta_{ij} \left(\frac{\kappa}{M^2}\right)^i (u M^2) ^j\,,
\end{align}
such that the indexes $i$ and $j$  indicate the degree of a polynomial expansion in $\kappa$ and $u$, respectively. %
We then systematically expand each $ \delta_{ij}$ in powers of $\chi_{1}$, $\chi_2$, and $\chi_{\rm eff}$. The resulting prefactors are all rational functions of $q$. We obtain:
\begingroup
\allowdisplaybreaks
\begin{widetext}
\begin{align}
\bar\delta &= 64 \frac{(1 + q)^6}{q} u^2 M^4\,,
\\
\delta_{56}&=\delta_{55}=\delta_{54}=\delta_{53}=\delta_{52}=0\,,
\\
\delta_{51}&=-1\,,
\\
\delta_{50}&=0 \,,
\\
\delta_{46}&=\delta_{45}=\delta_{44}=\delta_{43}=0 \,,
\\
\delta_{42}&=\frac{5-3 q}{(1+q)^4} \chi_1^2 -\frac{q^3 (3-5 q 
  ) }{(1+q)^4} \chi_2^2+\frac{q }{(1+q)^2} \chi_{\rm eff}^2 \,,
   \\
\delta_{41} &= \frac{5 }{2}\chi_{\rm eff}\,,
\\
\delta_{40} &= \frac{(1-q)^2}{16 q}\,,
\\
\delta_{36}&=\delta_{35}=\delta_{34}=0\,,
\\
\delta_{33} &= -\frac{10-12 q+3 q^2 }{(1+q)^8} \chi_{1}^4 +\frac{2 q^3 (6-11 q+6 q^2)
  }{(1+q)^8}  \chi_1^2 \chi_2^2-\frac{q^6
   (3-12 q+10 q^2)}{(1+q)^8}  \chi_{2}^4-\frac{2 q (2-q) }{(1+q)^6} \chi_1^2 \chi_{\rm eff}^2+\frac{2 q^4 (1-2 q) }{(1+q)^6}  \chi_2^2
   \chi_{\rm eff}^2\,,
\\
\delta_{32} &= -\frac{20-3 q-q^2 }{2 (1+q)^5} \chi_1^2 \chi_{\rm eff} +\frac{q^3 (1+3 q-20 q^2)
 }{2 (1+q)^5}   \chi_2^2 \chi_{\rm eff} -\frac{2 q
  }{(1+q)^2}  \chi_{\rm eff}^3\,,
\\
\delta_{31}&=-\frac{(1-q)^2 (1-5 q)  }{4 q
   (1+q)^4} \chi_1^2 +\frac{q^2(1-q)^2 (5-q)}{4
   (1+q)^4}  \chi_2^2 -\frac{2 (1+3 q+q^2) }{(1+q)^2} \chi_{\rm eff}^2\,,
\\
\delta_{30}&= -\frac{(1-q)^2}{8 q}  \chi_{\rm eff}\,,
\\
\delta_{26}&=\delta_{25}=0\,,
\\
\delta_{24} &= \frac{(1-q) (10-8 q+q^2) }{(1+q)^{12}}\chi_1^6 -\frac{9 q^3 (1-q)  (2-2 q+q^2)
   }{(1+q)^{12}}\chi_1^4 \chi_2^2
   +\frac{9 q^6
   (1-q)  (1-2 q+2 q^2) }{(1+q)^{12}} \chi_1^2
   \chi_2^4
        \notag \\ &
   -\frac{q^9  (1-q) (1-8 q+10
   q^2) }{(1+q)^{12}}\chi_2^6+\frac{q (6-6
   q+q^2) }{(1+q)^{10}}\chi_1^4 \chi_{\rm eff}^2 
   -\frac{2 q^4 (3-5 q+3 q^2)
   }{(1+q)^{10}}\chi_1^2 \chi_2^2 \chi_{\rm eff}^2+\frac{q^7 (1-6 q+6 q^2)
   }{(1+q)^{10}}\chi_2^4 \chi_{\rm eff}^2\,,
\\
\delta_{23} &= \frac{30-39 q+19 q^2-4 q^3 }{2 (1+q)^9} \chi_1^4
   \chi_{\rm eff} -\frac{3 q^3 (1+q^2)
  }{2
   (1+q)^8} \chi_1^2 \chi_2^2 \chi_{\rm eff} -\frac{q^6 (4-19 q+39 q^2-30 q^3)
  }{2 (1+q)^9}  \chi_2^4 \chi_{\rm eff} 
\notag  \\&
+\frac{2 q
   (3-q+q^2)}{(1+q)^7} \chi_1^2 \chi_{\rm eff}^3+\frac{2 q^4 (1-q+3 q^2)
 }{(1+q)^7}   \chi_2^2 \chi_{\rm eff}^3   \,,
\\
\delta_{22} &=\frac{(1-q)^2 (3-30 q-4 q^2)}{8 q
   (1+q)^8} \chi_1^4 -\frac{q^2 (1-q)^2  (15-29 q+15 q^2)
   }{4 (1+q)^8}\chi_1^2 \chi_2^2-\frac{q^5 (1-q)^2
    (4+30 q-3 q^2) }{8
   (1+q)^8}\chi_2^4
 \notag  \\&
   +\frac{12+2 q+11 q^2-q^3}{2 (1+q)^6} \chi_1^2 \chi_{\rm eff}^2-\frac{q^3 (1-11
   q-2 q^2-12 q^3) }{2 (1+q)^6} \chi_2^2 \chi_{\rm eff}^2 +\frac{q (1+4 q+q^2)
   }{(1+q)^4} \chi_{\rm eff}^4\,,
\\
\delta_{21} &= \frac{(1-q)^2 (3-23 q-4 q^2)}{8 q (1+q)^5}  \chi_1^2
   \chi_{\rm eff} -\frac{ q^2 (1-q)^2 (4+23
   q-3 q^2) }{8
   (1+q)^5} \chi_2^2 \chi_{\rm eff} +\frac{1+8 q+q^2 }{2
   (1+q)^2} \chi_{\rm eff}^3\,,
\\
\delta_{20} &= -\frac{(1-q)^4 }{16 q (1+q)^4} \chi_1^2 -\frac{q (1-q)^4 
   }{16 (1+q)^4} \chi_2^2 +\frac{(1-q)^2 (1+4
   q+q^2)}{16 q (1+q)^2}  \chi_{\rm eff}^2\,,
\\
\delta_{16} & =0\,,
\\
\delta_{15} &=-\frac{(1-q)^2  (5-2 q) }{(1+q)^{16}}\chi_1^8 +\frac{2 q^3
   (1-q)^2  (6-q+q^2)  }{(1+q)^{16}}\chi_1^6\chi_2^2 -\frac{9 q^6 (1-q)^2  (1+q^2)
   }{(1+q)^{16}}\chi_1^4 \chi_2^4
 \notag  \\&
 +\frac{2 q^9 
   (1-q)^2 (1-q+6 q^2) }{(1+q)^{16}}\chi_1^2
   \chi_2^6 +\frac{q^{13} (1-q)^2 (2-5 q)  
   }{(1+q)^{16}}\chi_2^8 -\frac{2 q  (1-q) (2-q) 
  }{(1+q)^{14}} \chi_1^6 \chi_{\rm eff}^2
 \notag  \\&
 +\frac{2 q^4
   (1-q)  (3-q+q^2) }{(1+q)^{14}} \chi_1^4 \chi_2^2 \chi_{\rm eff}^2 -\frac{2 q^7 (1-q) 
   (1-q+3 q^2)}{(1+q)^{14}} \chi_1^2 \chi_2^4
   \chi_{\rm eff}^2 -\frac{2 q^{11}(1-q)  (1-2 q)  
 }{(1+q)^{14}}   \chi_2^6 \chi_{\rm eff}^2\,,
\\
\delta_{14} &= 
-\frac{(1-q) (20-29 q+12 q^2) }{2 (1+q)^{13}} \chi_1^6
   \chi_{\rm eff} +\frac{q^3(1-q) (3+14
   q^2-8 q^3)}{2 (1+q)^{13}} \chi_1^4 \chi_2^2
   \chi_{\rm eff} +\frac{q^6 (1-q) (8-14
   q-3 q^3)}{2 (1+q)^{13}}  \chi_1^2 \chi_2^4 \chi_{\rm eff} 
 \notag  \\&
   +\frac{q^{10} (1\!-\!q)  (12\!-\!29 q\!+\!20
   q^2) }{2
   (1+q)^{13}}\chi_2^6 \chi_{\rm eff} -\frac{2 q (3\!-\!5 q\!+\!3 q^2) }{(1+q)^{11}}\chi_1^4 \chi_{\rm eff}^3 -\frac{2q^4 (1\!-\!2 q) (2\!-\!q)
    }{(1+q)^{10}}\chi_1^2 \chi_2^2 \chi_{\rm eff}^3 -\frac{2 q^8 (3\!-\!5 q\!+\!3 q^2)
   }{(1+q)^{11}} \chi_2^4 \chi_{\rm eff}^3\,,
\\
\delta_{13} &= 
-\frac{(1\!-\!q)^2 (1-15 q-4 q^2)}{4 q
   (1+q)^{12}} \chi_1^6 +\frac{q^2 (1\!-\!q)^2  (15-55 q+18 q^2+4
   q^3) }{4
   (1+q)^{12}}\chi_1^4 \chi_2^2 +\frac{q^5 (1\!-\!q)^2  (4+18 q-55 q^2+15
   q^3)}{4
   (1+q)^{12}}  \chi_1^2 \chi_2^4 
    \notag  \\&
   +\frac{ q^9 (1-q)^2 (4+15 q-q^2)
 }{4 (1+q)^{12}}   \chi_2^6 -\frac{6-16 q+18 q^2-5
   q^3}{(1+q)^{10}}  \chi_1^4 \chi_{\rm eff}^2 +\frac{q^3 (1-8 q+20 q^2-8
   q^3+q^4)}{(1+q)^{10}} \chi_1^2 \chi_2^2 \chi_{\rm eff}^2 
   \notag \\ &
   +\frac{q^7 (5-18 q+16 q^2-6
   q^3) }{(1+q)^{10}} \chi_2^4 \chi_{\rm eff}^2 -\frac{2 q (1-q+3 q^2)
  }{(1+q)^8} \chi_1^2 \chi_{\rm eff}^4 -\frac{2 q^5
   (3-q+q^2) }{(1+q)^8} \chi_2^2 \chi_{\rm eff}^4\,,
\\
\delta_{12} &= -\frac{(1-q)^2 (3-49 q-16 q^2)}{8 q (1+q)^9}  \chi_1^4
   \chi_{\rm eff} +\frac{q^2(1-q)^2 (4-7
   q+4 q^2)}{4 (1+q)^8} \chi_1^2 \chi_2^2 \chi_{\rm eff} +\frac{q^6 (1-q)^2  (16+49 q-3 q^2)
 }{8
   (1+q)^9}  \chi_2^4 \chi_{\rm eff}
   \notag \\ &   
   -\frac{2-5 q+19 q^2 }{2 (1+q)^7}\chi_1^2
   \chi_{\rm eff}^3 -\frac{q^5 (19-5 q+2
   q^2) }{2
   (1+q)^{7}}\chi_2^2 \chi_{\rm eff}^3 -\frac{2 q^2 }{(1+q)^4} \chi_{\rm eff}^5 \,,
\\
\delta_{11} &= 
\frac{(1-q)^4}{8 q (1+q)^8}  \chi_1^4 +\frac{q (1-q)^4 
   (1-12 q+q^2) }{8
   (1+q)^8} \chi_1^2 \chi_2^2 +\frac{q^5 (1-q)^4 }{8
   (1+q)^8}  \chi_2^4 -\frac{(1-q)^2 (1-18 q-7 q^2) }{8 q (1+q)^6}\chi_1^2 \chi_{\rm eff}^2 
     \notag \\ &   
   +\frac{q^3 (1-q)^2    (7+18 q-q^2)}{8 (1+q)^6} \chi_2^2 \chi_{\rm eff}^2 -\frac{q (1+3 q+q^2) }{(1+q)^4} \chi_{\rm eff}^4\,,
   \\
\delta_{10} & = 
\frac{(1-q)^4 }{8 q
   (1+q)^5} \chi_1^2 \chi_{\rm eff} +\frac{q^2 (1-q)^4  }{8 (1+q)^5}\chi_2^2 \chi_{\rm eff} -\frac{(1-q)^2}{8
   (1+q)^2}  \chi_{\rm eff}^3\,,
\end{align}
\begin{align}
\delta_{06} &=
\frac{(1-q)^3}{(1+q)^{20}} \chi_1^{10} -\frac{q^3 (1-q)^3 
   (3+2 q) }{(1+q)^{20}}\chi_1^8 \chi_2^2+\frac{q^6 (1-q)^3  (3+6 q+q^2)
  }{(1+q)^{20}} \chi_1^6 \chi_2^4 -\frac{ q^9 (1-q)^3
   (1+6 q+3 q^2)}{(1+q)^{20}} \chi_1^4 \chi_2^6 
   \notag \\ &
   +\frac{q^{13} (1-q)^3  (2+3 q)}{(1+q)^{20}} \chi_1^2 \chi_2^8 -\frac{ q^{17} (1-q)^3
  }{(1+q)^{20}} \chi_2^{10} +\frac{q (1-q)^2 }{(1+q)^{18}} \chi_1^8 \chi_{\rm eff}^2 -\frac{2 q^4 (1-q)^2 
  }{(1+q)^{17}} \chi_1^6 \chi_2^2 \chi_{\rm eff}^2 
     \notag \\ &
  +\frac{q^7  (1-q)^2 (1+4 q+q^2)
  }{(1+q)^{18}} \chi_1^4 \chi_2^4 \chi_{\rm eff}^2 -\frac{2 q^{11}  (1-q)^2 }{(1+q)^{17}} \chi_1^2
   \chi_2^6 \chi_{\rm eff}^2 +\frac{q^{15} (1-q)^2 }{(1+q)^{18}}  \chi_2^8
   \chi_{\rm eff}^2\,,
\\
\delta_{05} &=
\frac{(1-q)^2 (5-8 q) }{2
   (1+q)^{17}} \chi_1^8 \chi_{\rm eff} -\frac{q^3(1-q)^2  (1-4 q)   (1+3 q) }{2
   (1+q)^{17}} \chi_1^6 \chi_2^2 \chi_{\rm eff} -\frac{q^6 (1-q)^2  (4+q+4 q^2)
  }{2
   (1+q)^{16}} \chi_1^4 \chi_2^4 \chi_{\rm eff} 
\notag \\ &
+\frac{q^{10} (1-q)^2 (4-q)  (3+q)}{2 (1+q)^{17}}  \chi_1^2
   \chi_2^6 \chi_{\rm eff} -\frac{q^{14} (1-q)^2 (8-5
   q)  }{2
   (1+q)^{17}}\chi_2^8 \chi_{\rm eff} +\frac{2  q (1-q) (1-2 q)}{(1+q)^{15}} \chi_1^6
   \chi_{\rm eff}^3 
   \notag \\ &
   +\frac{2 q^4   (1-q)(1-q+3
   q^2) }{(1+q)^{15}} \chi_1^4 \chi_2^2 \chi_{\rm eff}^3 -\frac{2  q^8 (1-q) (3-q+q^2)
  }{(1+q)^{15}} \chi_1^2 \chi_2^4 \chi_{\rm eff}^3 +\frac{2 q^{12} (1-q) (2-q)  }{(1+q)^{15}} \chi_2^6 \chi_{\rm eff}^3\,,
\\
\delta_{04} &=
\frac{(1\!-\!q)^2 (1-20 q-8 q^2)}{16 q
   (1+q)^{16}}  \chi_1^8 -\frac{q^2 (1\!-\!q)^2  (5-27 q+3 q^2-8 q^3)
  }{4
   (1+q)^{16}} \chi_1^6 \chi_2^2 
      -\frac{q^5 (1-q)^2  (4+6 q+61 q^2+6 q^3+4
   q^4)}{8
   (1+q)^{16}} \chi_1^4 \chi_2^4 
       \notag \\ &
   +\frac{q^9 (1-q)^2  (8-3 q+27 q^2-5 q^3)
   }{4
   (1+q)^{16}}\chi_1^2 \chi_2^6 
      -\frac{q^{13} (1-q)^2  (8+20 q-q^2)
  }{16 (1+q)^{16}} \chi_2^8 +\frac{(1-q) (4-18 q+11
   q^2) }{2
   (1+q)^{14}} \chi_1^6 \chi_{\rm eff}^2
          \notag \\ &
    -\frac{ q^3 (1-q) (1-4 q-14 q^2+8 q^3)
   }{2
   (1+q)^{14}}\chi_1^4 \chi_2^2 \chi_{\rm eff}^2 +\frac{ q^7  (1-q)(8-14 q-4 q^2+q^3)
  }{2
   (1+q)^{14}} \chi_1^2 \chi_2^4 \chi_{\rm eff}^2 
   \notag \\ &
   -\frac{q^{11}  (1\!-\!q) (11-18 q+4 q^2)
   }{2 (1+q)^{14}} \chi_2^6 \chi_{\rm eff}^2
   +\frac{q
   (1-6 q+6 q^2)}{(1+q)^{12}}  \chi_1^4 \chi_{\rm eff}^4 -\frac{2 q^5 (3-5 q+3 q^2)
  }{(1+q)^{12}} \chi_1^2 \chi_2^2 \chi_{\rm eff}^4 +\frac{q^9 (6-6 q+q^2)
}{(1+q)^{12}}   \chi_2^4 \chi_{\rm eff}^4 \,,
   \\
\delta_{03} &=
\frac{(1-q)^2 (1-25 q-12 q^2) }{8 q (1+q)^{13}} \chi_1^6
   \chi_{\rm eff} -\frac{q^2 (1-q)^2 
   (4-29 q+21 q^2-32 q^3) }{8
   (1+q)^{13}} \chi_1^4
   \chi_2^2 \chi_{\rm eff}
\notag \\ &   
   +\frac{ q^6  (1-q)^2(32-21 q+29 q^2-4
   q^3)}{8 (1+q)^{13}}  \chi_1^2 \chi_2^4 \chi_{\rm eff} -\frac{q^{10} (1-q)^2  (12+25
   q-q^2) }{8
   (1+q)^{13}}\chi_2^6 \chi_{\rm eff} +\frac{1-14 q+20 q^2-5 q^3 }{2 (1+q)^{11}}\chi_1^4 \chi_{\rm eff}^3 
 \notag \\ &     
   -\frac{q^4 (2-3
   q+2 q^2) }{(1+q)^{10}}\chi_1^2 \chi_2^2 \chi_{\rm eff}^3 -\frac{q^8 (5-20 q+14 q^2-q^3)
  }{2 (1+q)^{11}} \chi_2^4 \chi_{\rm eff}^3 -\frac{2
 q^2  (1-2 q) }{(1+q)^9} \chi_1^2 \chi_{\rm eff}^5 +\frac{2  q^6 (2-q)}{(1+q)^9}  \chi_2^2
   \chi_{\rm eff}^5\,,
 \\
\delta_{02} &=
-\frac{(1-q)^4}{16 q (1+q)^{12}}  \chi_1^6 -\frac{q  (1-q)^4
   (1-24 q-10 q^2)}{16 (1+q)^{12}} \chi_1^4 \chi_2^2 +\frac{q^5 (1-q)^4  (10+24
   q-q^2) }{16
   (1+q)^{12}}\chi_1^2 \chi_2^4 -\frac{q^9 (1-q)^4 }{16
   (1+q)^{12}} \chi_2^6 
\notag \\ &
   +\frac{(1-q)^2 (1-40 q-23 q^2)
   }{16 q
   (1+q)^{10}}\chi_1^4 \chi_{\rm eff}^2 +\frac{q^3 (1-q)^2  (11-24 q+11 q^2)
   }{8
   (1+q)^{10}}\chi_1^2 \chi_2^2 \chi_{\rm eff}^2 -\frac{q^7 (1-q)^2  (23+40 q-q^2)
  }{16 (1+q)^{10}} \chi_2^4 \chi_{\rm eff}^2 
  \notag \\ &
  -\frac{q
   (4-7 q-q^2) }{2 (1+q)^8} \chi_1^2 \chi_{\rm eff}^4+\frac{q^5 (1+7 q-4 q^2)
  }{2 (1+q)^8}  \chi_2^2 \chi_{\rm eff}^4 +\frac{q^3
   }{(1+q)^6}\chi_{\rm eff}^6 \,,
\\
\delta_{01} & = -\frac{(1-q)^4 }{8 q
   (1+q)^9}\chi_1^4 \chi_{\rm eff} +\frac{5 q^2 (1-q)^4  }{8 (1+q)^8}\chi_1^2 \chi_2^2 \chi_{\rm eff} -\frac{q^6 (1-q)^4 
   }{8 (1+q)^9}\chi_2^4 \chi_{\rm eff} -\frac{(1-q)^2
   (5+3 q)}{8
   (1+q)^7} \chi_1^2 \chi_{\rm eff}^3
     \notag \\ &
    -\frac{ q^4  (1-q)^2(3+5 q)}{8 (1+q)^7} \chi_2^2
   \chi_{\rm eff}^3 +\frac{q^2 }{2 (1+q)^4}\chi_{\rm eff}^5 \,,
\\
\delta_{00} & = 
\frac{q(1-q)^6 }{16
   (1+q)^8} \chi_1^2 \chi_2^2-\frac{(1-q)^4 }{16 q (1+q)^6} \chi_1^2 \chi_{\rm eff}^2 -\frac{q^3 (1-q)^4  }{16 (1+q)^6}\chi_2^2
   \chi_{\rm eff}^2 +\frac{q (1-q)^2 }{16 (1+q)^4}  \chi_{\rm eff}^4\,.
 \end{align}
\end{widetext}%
\endgroup

The coefficients entering the expansion  
\begin{align}
\chi_{\rm p}^2 & = \bar\lambda \sum_{i=0}^2  \lambda_i\delta\chi^i
\end{align}
of Eq.~(\ref{chip2poly})  are %
\begin{align}
\bar\lambda &=\frac{1+q}{4 q (4+3q)^2 u M^2}\,,
\end{align}
\begin{align}
\lambda_2 &= - q (1-q)^2 (1+q) u M^2\,,
\\
\lambda_1 &= - 2 (1-q) (1+q)^2 \left[ (4+3q) (3+4q) + 7 q  \chi_{\rm eff} u M^2\right] ,
\\
\lambda_0 & = 2 (1+q)^3 (4+3q) (3+4q) \left(2 \frac{\kappa}{M^2} -\chi_{\rm eff} \right) 
\notag \\
&-u M^2 \Big\{12 (1-q) \left[(4+3q)
   \chi_1^2-q^3 (3+4q) \chi_2^2\right]
  \notag \\
   &+49 q (1+q)^3 \chi_{\rm eff}^2\Big\}\,.
\end{align}

\section{Polynomial root finding}
\label{rootsalg}

Accurately determining roots of polynomials is a crucial ingredient for the successful implementation of our formalism and dominates the overall computational cost (Sec.~\ref{secprof}). Most notably, this is needed to compute $\delta\chi_{\pm,3}$ and $\kappa_\pm$. 
We find that the implementation readily available in \texttt{numpy.roots}  \cite{2020Natur.585..357H} provides the necessary accuracy. However, it is not vectorized and can only handle one root-finding problem at a time. %

The mathematical problem is the following: we wish to solve $N$ equations of the kind
\begin{align}
\left\{ \sum_{j=0}^M c_{ij} x^{M-j} = 0 \right\} _{i=1}^{N}\,,
\end{align}
where $M$ is (larger than) the largest degree of all the equations. The input is provided by the $N\times M$ matrix $C$ with elements $c_{ij}$.

Array vectorization can be achieved by appropriately inserting
{new array axes}
in the \texttt{numpy.roots} code \cite{stackoverflow}. This, however, requires all equations in the array to be of the same degree as values $c_{i0}=0$ cannot be accommodated by the adopted companion-matrix algorithm~\cite{horn2012matrix}. The public version of \texttt{numpy.roots} addresses this issue by stripping all trailing zeros from the input rank-1 array $c_{1j}$ before performing the required linear-algebra operations. This is not viable when considering $N>1$ equations because the number of trailing zeros could be different in each row of the coefficient matrix $C$. For instance, this is the case when considering binaries with both $q<1$ and $q=1$, such that some of the resulting equations $\Sigma(\delta\chi)=0$ are cubic and others are quadratic; cf. Sec.~\ref{notable}.

We solve this issue by identifying the number of trailing zeros in each row and applying a suitable permutation to the element of that row such that those zeros end up in the last columns. The resulting equations are all of the same degree and present a number of additional null roots equal to the number of trailing zeros in the original problem. These spurious solutions can then be easily filtered out or masked.

This is best explained with an example. Consider the
set of equations
\begin{align}
\label{polyex}
\begin{dcases}
x^3- 6 x^2 + 11 x -6 =0\,,\\
x^2 - 3 x + 2 = 0\,,\\
x^4 - 10 x^3 + 35 x^2- 50 x + 24  = 0\,,  \\
\end{dcases}
\end{align}
with solutions
\begin{align}
\label{polysolve}
\begin{dcases}
x= 1,2, 3\,,\\
x = 1,2\,, \\
x=1,2,3,4\,.  \\
\end{dcases}
\end{align}
The coefficient matrix  from Eqs.~(\ref{polyex}) is
\begin{align}
	C &=
	\begin{pmatrix}
		0 & 1 & -6 & 11 & -6 \\
		0 & 0 & 1& -3 & 2 \\
		1 & -10 & 35 & -50 & 24 \\
	\end{pmatrix} \,.
\end{align}
The number of trailing zeros in each row is $n_{t0}=(1,2,0)$. We cycle the coefficient of each row a number of times given by $n_{t0}$. This results in a modified coefficient matrix,
\begin{align}
	C' &=
	\begin{pmatrix}
		1 & -6 & 11 & -6 & 0  \\
		1& -3 & 2 & 0 & 0  \\
		1 & -10 & 35 & -50 & 24 \\
	\end{pmatrix} \, ,
\end{align}
that can be easily digested by the vectorized version of \texttt{numpy.roots} \cite{stackoverflow}. The corresponding equations are 
\begin{align}
\label{polyex2}
\begin{dcases}
x'^4- 6 x'^3 + 11 x'^2 -6x' = x' (x'^3- 6 x'^2 + 11 x' -6) =0 \,,\\
x'^4 - 3 x'^3 + 2x'^2 =  x'^2( x'^2 - 3 x' + 2) = 0\,,\\
x'^4 - 10 x'^3 + 35 x'^2- 50 x' + 24  = 0\,,  \\
\end{dcases}
\end{align}
with solutions
\begin{align}
\label{newsols}
\begin{dcases}
x'= 0, 1,2, 3\,,\\
x' = 0,0, 1,2\,, \\
x'=1,2,3,4\,.  \\
\end{dcases}
\end{align}
Masking a number of zeros equal to $n_{t0}=(1,2,0)$ in each set returns the solutions of the original problem; cf. Eq.~(\ref{polysolve}).

In short, our algorithm provides an array-compatible generalization of \texttt{numpy.roots} that can handle multiple equations of different degrees.

\section{Useful integrals}
\label{appintegrals}
In this Appendix, we report some standard integrals that are used in this paper. 

Let us first recall the definition of the elliptic integrals in their Legendre form; see Ref.~\cite{1965hmfw.book.....A} for a complete introduction. Given $0\leq\varphi\leq  \pi/2$ and $0\leq m \leq 1$, one defines the following special functions.
\begin{itemize}
\item Incomplete elliptic integral of the first kind:
\begin{align}
F(\varphi,m) = \int_0^\varphi \frac{\dd \theta}{\sqrt{1- m \sin^2 \theta}}\,.
\end{align}
\item Complete elliptic integral of the first kind:
\begin{align}
K(m) = F\left(\varphi = \frac{\pi}{2},m\right)\,.
\end{align}
\item Incomplete elliptic integral of the first kind:
\begin{align}
E(\varphi,m) = \int_0^\varphi  \sqrt{1- m \sin^2 \theta}~\dd\theta\,.
\end{align}
\item Complete elliptic integral of the second kind:
\begin{align}
E(m) = E\left(\varphi = \frac{\pi}{2},m\right)\,.
\end{align}
\item Incomplete elliptic integral of the third kind:
\begin{align}
\Pi(n,\varphi,m) =\int_0^\varphi \frac{\dd\theta}{(1-n\sin^2\theta)\sqrt{1- m \sin^2 \theta}}\,.
\end{align}
\item Complete elliptic integral of the third kind:
\begin{align}
\Pi(n,m)= \Pi\left(n,\varphi = \frac{\pi}{2},m\right)\,.
\end{align}
\end{itemize}

Let us assume $a< x < b \leq c$. The following integrals involving the cubic function $(x-a)(b-x)(c-x)>0$ can be reduced as follows:
\begingroup
\allowdisplaybreaks
\begin{align}
&\int \frac{1}{\sqrt{(x-a)(b-x)(c-x)}} \,\dd x\notag
\\ &\qquad =\frac{2}{\sqrt{c-a}}{F\left(\arcsin\sqrt\frac{x-a}{b-a}, \frac{b-a}{c-a}\right)}\,,\\
&\int_a^b \frac{1}{\sqrt{(a-x)(b-x)(c-x)}} \,\dd x= \frac{2}{\sqrt{c-a}} K\left(\frac{b-a}{c-a}\right)\,,
\\
&\int \frac{x}{\sqrt{(x-a)(b-x)(c-x)}} \,\dd x\notag
\\ &\qquad 
=\frac{2}{\sqrt{c-a}} \bigg[c F\left(\arcsin\sqrt\frac{x-a}{b-a},\frac{b-a}{c-a}\right)
 \notag  \\&\qquad-(c-a) E\left(\arcsin\sqrt\frac{x-a}{b-a},\frac{b-a}{c-a}\right)\bigg]\,,
 \end{align}
\pagebreak

\begin{align}
&\int_a^b \frac{x}{\sqrt{(x-a)(b-x)(c-x)}}\, \dd x 
\notag
\\
&\qquad= \frac{2}{\sqrt{c-a}} \left[c K\left(\frac{b-a}{c-a}\right)-(c-a)E\left(\frac{b-a}{c-a}\right)\right]\,,
\\
&\int \frac{x^2}{\sqrt{(x-a)(b-x)(c-x)}} \,\dd x\notag
\\ &\qquad 
=\frac{2}{3
   \sqrt{c-a}} \bigg\{ \sqrt{(c-a) (x-a) (b-x) (c-x)}
   \notag \\&\qquad 
 +[ b (c-a)+c (a+2 c) ] F\left(\arcsin
 \sqrt{\frac{x-a}{b-a}},\frac{b-a}{c-a}\right)
   \notag \\&\qquad 
 -2 (c-a) (a+b+c)
   E\left(\arcsin
   \sqrt{\frac{x-a}{b-a}},\frac{b-a}{c-a}\right) \bigg\}\,,
 \\
&\int_a^b \frac{x^2}{\sqrt{(x-a)(b-x)(c-x)}} \,\dd x 
\notag
\\ &\qquad 
=\frac{2}{3
   \sqrt{c-a}} \bigg\{  
[ b (c-a)+c (a+2 c) ] K\left(\frac{b-a}{c-a}\right)
   \notag \\&\qquad 
 -2 (c-a) (a+b+c)
   E\left(\frac{b-a}{c-a}\right) \bigg\}\,,
 \\
&\int \frac{1}{(k-x)\sqrt{(x-a)(b-x)(c-x)}} \,\dd x\notag
\\ &\qquad =\frac{2}{\sqrt{c-a}(k-a)}  \Pi\left(\frac{b-a}{k-a},\arcsin\sqrt\frac{x-a}{b-a},\frac{b-a}{c-a}\right)\,,
\\
&\int_a^b \frac{1}{(k-x)\sqrt{(x-a)(b-x)(c-x)}} \,\dd x\notag
\\ &\qquad = \frac{2}{\sqrt{c-a}(k-a)}  \Pi\left(\frac{b-a}{k-a},\frac{b-a}{c-a}\right)\,.
\end{align}
\endgroup

\bibliography{fastprecession}

\end{document}